\documentclass[journal, 12pt,onecolumn, draftcls]{IEEEtran}
\usepackage{graphicx}
  \usepackage{amsmath,graphicx}
  \usepackage{stfloats}
      \usepackage{mathdots}
  \usepackage{amssymb,bm}
  \usepackage{amsfonts}
  \usepackage{amscd}
  \usepackage{cite}
  \usepackage[dvips]{color}
  \usepackage{epsfig}
  \usepackage{dsfont}
  \usepackage{textcomp}
  \usepackage{hyperref}
  \usepackage{etoolbox}
  \usepackage{verbatim}
  \usepackage{lettrine}
  \usepackage{algorithm}
  \usepackage{algorithmicx}
  \usepackage{algpseudocode}
  \hypersetup{linktocpage} 
  \hypersetup{
    colorlinks,
    citecolor=blue,
    filecolor=black,
    linkcolor=blue,
    urlcolor=black
}
 \author{
\IEEEauthorblockN{{Rohit Budhiraja, Karthik KS and Bhaskar Ramamurthi\\}
\thanks{A part of this work was presented in ICC-2013. The authors are with the Dept. of Electrical Engineering, Indian Institute of Technology Madras, Chennai-600036, India (email:\{ee11d021,ee08d024\}@ee.iitm.ac.in, bhaskar@iitm.ac.in).}
}
 }
\newtheorem{remark}{Remark}
\newtheorem{theorem}{Theorem}[section]
\newtheorem{lemma}[theorem]{Lemma}
\algnewcommand{\algorithmicgoto}{\textbf{go to}}%
\algnewcommand{\Goto}[1]{\algorithmicgoto~\ref{#1}}%

\newtoggle{SINGLE_COL}
\toggletrue{SINGLE_COL}
\newtoggle{DOUBLE_COL}
\toggletrue{DOUBLE_COL}
\togglefalse{DOUBLE_COL}

\begin{document}
\title{Linear Precoders for Non-Regenerative Asymmetric Two-way Relaying in Cellular Systems}
\maketitle
\begin{abstract}
Two-way relaying (TWR) reduces the spectral-efficiency loss caused in conventional half-duplex relaying. TWR is possible when two nodes exchange data simultaneously through a relay. In cellular systems, data exchange between base station (BS) and users is usually not simultaneous e.g., a user (TUE) has uplink data to transmit during multiple access (MAC) phase, but does not have downlink data to receive during broadcast (BC) phase. This non-simultaneous data exchange will reduce TWR to spectrally-inefficient conventional half-duplex relaying. With infrastructure relays, where multiple users communicate through a relay, a new transmission protocol is proposed to recover the spectral loss. The BC phase following the MAC phase of TUE is now used by the relay to transmit downlink data to another user (RUE). RUE will not be able to cancel the back-propagating interference. A structured precoder is designed at the multi-antenna relay to cancel this interference. With multiple-input multiple-output (MIMO) nodes, the proposed precoder also triangulates the compound MAC and BC phase MIMO channels. The channel triangulation reduces the weighted sum-rate optimization to power allocation problem, which is then cast as a geometric program. Simulation results illustrate the effectiveness of the proposed protocol over conventional solutions.
\end{abstract}
\section{Introduction}
Cooperative communication is a promising technique which can lead to significant performance gains in the wireless systems including coverage extension and throughput enhancement. An example of cooperative communication is the conventional half-duplex two-hop one-way relaying \cite{owr_opt_pre_ref,owr_opt_pre_ref1,gp_owr_ref}. The half-duplex constraint in a relay station (RS) prevents it from receiving and transmitting simultaneously on the same channel. Communication through a conventional relay therefore requires four channel uses for bi-directional communication between two nodes, which is twice the number of channel uses required when two nodes communicate directly without a relay. TWR has been proposed to reduce this spectral-efficiency loss\cite{rankov2007spectral,pnc_ref0,relay_csi_sup_nw_code, twr_opt_pre_ref0, eurasip_af_mimo_ref, twr_opt_pre_ref5, twr_opt_pre_ref2, phy_nw_coding_ref0,twr_opt_pre_ref6,lau_mu_twr_ref}. 

During the first channel use in TWR, two source nodes simultaneously transmit their data signals to the relay. In the second channel use, relay broadcasts a function of the sum-signal received earlier during the first phase. The first and the second channel use are commonly known as the  multiple-access (MAC) and the broadcast (BC) phases, respectively. The key idea in TWR is that both source nodes can subtract the \textit{self-interference} from the sum-signal received in the BC phase, provided the required channel state information (CSI) is available. \textit{Self-interference}, also called \textit{back-propagating interference} (BI) in \cite{rankov2007spectral}, refers to the self-data of a node, transmitted back to the node by the relay. BI cancellation ensures an interference-free channel for both the nodes. TWR thus requires two channel uses for bi-directional data exchange as in direct communication, and recovers the loss in spectral-efficiency.

The underlying assumption in TWR is that two source nodes always have data to exchange simultaneously. However, in a cellular system, a user (UE) might have downlink data to receive from the BS but might not have uplink data to transmit to the BS at the same time \cite{infra_relay_ref0}. This practical constraint will reduce  simultaneous bi-directional data exchange to unidirectional data flow between the BS and a UE. With uni-directional data flow, TWR has the same inefficiency as the conventional one-way relaying. 

\begin{figure*}[htp]
    \begin{center}
    \includegraphics[width=5.5in]{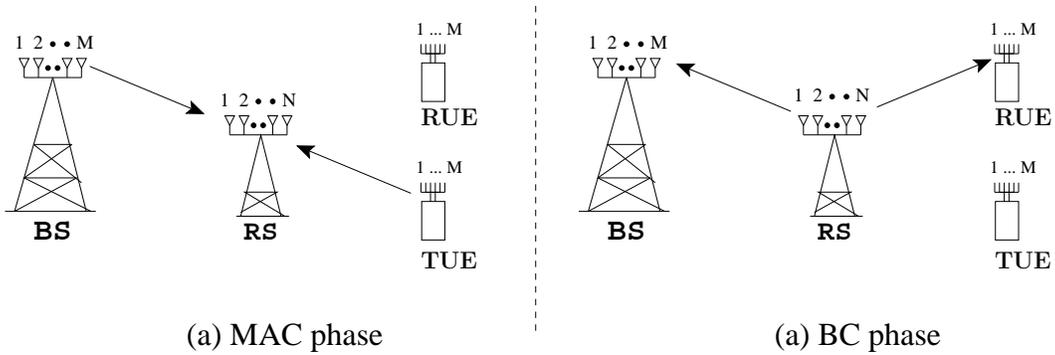} 
    \caption{\small Illustration of asymmetric TWR. During MAC phase, BS transmits data to be sent to the RUE, while TUE transmits data to be sent to the BS. During BC phase, the relay transmits a function of the sum-signal received during the MAC phase to the BS and RUE.}
    \label{2WayRelaying3Node}
    \end{center}
    \end{figure*}

Cellular systems are multi-user systems. Infrastructure relays \cite{infra_relay_ref0,infra_relay_ref1,infra_relay_ref2} have been proposed in the cellular systems to enable a BS serve multiple users through a relay. Now, consider a UE (say, RUE) that is downloading data from a network (in the downlink), but has no data to upload. Due to multiple users in the system, it is possible to  find another UE (say, TUE) which wants to transmit data to the BS with a high probability. We exploit this multi-user feature and propose a novel TWR transmission protocol to recover the spectral loss caused due to non-simultaneous data flow. We propose that, during MAC phase, BS transmits data to be communicated to the RUE, while TUE transmits data to be communicated to the BS as shown in Fig. \ref{2WayRelaying3Node}(a). Both these signals are received by the relay.  During BC phase, the relay will transmit a function of the sum-signal received earlier during the MAC phase to the BS and RUE, as shown in Fig. \ref{2WayRelaying3Node}(b). The new protocol enables exchange of two data units over two channel uses by re-establishing the bi-directional flow of traffic on either directions of the relay, resulting in a more efficient channel use. 

The two-way relaying now becomes \textit {asymmetric}, as two different UEs are served during the MAC and BC phases. However, due to this asymmetry, only BS can perform the BI cancellation. RUE will not be able to cancel the BI in the absence of necessary \textit{side-information}.\footnote{In this work, we assume that RUE cannot overhear the MAC phase transmission of TUE.} In the models considered in the existing literature \cite{rankov2007spectral,relay_csi_sup_nw_code, twr_opt_pre_ref0, twr_opt_pre_ref5, twr_opt_pre_ref2, phy_nw_coding_ref0,twr_opt_pre_ref6}, it is assumed that the data exchange is simultaneous, or nodes have the necessary side information to cancel the BI. In this paper, we extend the scope of TWR by incorporating the non-simultaneous downlink and uplink data flows observed in the cellular systems. 

In the symmetric TWR\footnote{In context of proposed asymmetric TWR, conventional TWR is referred as symmetric TWR in this paper.}, as the BI can be completely cancelled by the receiving nodes, the precoder design is done exclusively to optimize a desired figure of merit e.g., minimize mean square error (MSE) or maximize weighted sum-rate \cite{twr_opt_pre_ref1,twr_opt_pre_ref2,twr_opt_pre_ref5}. On the other hand, for the addressed communication scenario, RUE will observe poor signal-to-interference-plus-noise ratio (SINR) in the presence of BI. It is therefore crucial to mitigate the asymmetric BI observed by the RUE to improve its SINR before optimizing any figure of merit. 

The current research has demonstrated the tremendous performance benefits of using multiple-input multiple-output (MIMO) nodes in the conventional one-way relaying and symmetric TWR  channels \cite{owr_opt_pre_ref,gp_owr_ref,relay_csi_sup_nw_code, twr_opt_pre_ref0, eurasip_af_mimo_ref, twr_opt_pre_ref5,twr_opt_pre_ref2}. The system model in the present work also assumes that all the nodes are equipped with multiple antennas. The BS and TUE have $M$ antennas and transmit $M$ independent data streams during the MAC phase. During BC phase, RUE will require a minimum of $2M$ antennas; $M$ antennas to suppress the BI and additional $M$ antennas to decode its desired data \cite{winter_int_can}. This is a prohibitive requirement for a UE, as the number of antennas used at the UE is typically small due to practical form-factor constraint \cite{ue_size_res_ref1}. Another solution to handle the BI problem is to restrict BS and TUE to transmit only $M/2$ streams during the MAC phase. RUE will now require only $M$ antennas to decode its $M/2$ streams. But this artificial restriction results in the under-utilization of available spatial resources, as the number of transmit streams reduces by a factor of half. Asymmetric TWR leads to a situation where communication between three nodes is possible either by satisfying the physically-limiting constraint of using $\ge 2M$ antennas at the RUE, or by sacrificing the available spatial resources.

The main challenge for the asymmetric TWR is to ensure that the signal received by the RUE is free from BI. This work aims to address this problem and designs a linear  precoder at the infrastructure relay to completely cancel the BI. The infrastructure relays do not have form-factor constraints unlike a UE \cite{infra_relay_ref0,infra_relay_ref1}. This precoder enables the BS and TUE to transmit $M$ streams during the MAC phase with RUE requiring only $M$ antennas to decode its desired data. The proposed precoder thus results in the full use of available spatial resources and transfers the complexity of cancelling the BI from the RUE to the relay. Furthermore, the proposed precoder also triangulates the MIMO MAC- and BC-phase channels. The channel-triangularization simplifies the RUE and BS receiver design considerably. 

In a cellular network, the quality of service (QoS) requirements normally lead to higher downlink data-rate than the uplink. The sum downlink-plus-uplink-rate maximization is therefore inappropriate in cellular scenario \cite{infra_relay_ref0}. For the asymmetric TWR protocol proposed for cellular scenario, it is important to maximize the weighted downlink-plus-uplink sum-rate instead. The problem of weighted sum-rate (WSR) maximization at the relay for asymmetric TWR is also addressed in this work. Due to channel-triangularization, WSR optimization is reduced to global power allocation problem at the relay. With the proposed precoder structure, WSR maximization will enable the relay to assign different priorities to each of the $2M$ downlink-plus-uplink streams to satisfy their respective QoS requirements. 

\textbf{Related work}: In \cite{murch_overhearing}, the authors propose a three-slot protocol for multi-user relaying and make an assumption that RUE can overhear and decode TUE  \textit{without any errors}. This is a strong assumption and is usually difficult to ensure in practice in cellular systems. In this work, we propose a two-slot protocol, and do not assume overhearing among UEs. 

Model in the present work is also different from the asymmetric data-rate model in \cite{asym_ref5,asym_ref6,asym_ref0,asym_ref1,asym_ref2}, where users exchange different amounts of data through a two-way relay. Moreover, \cite{asym_ref5,asym_ref6,asym_ref0,asym_ref1,asym_ref2} consider only one UE and not multiple UEs served by the relay, and asymmetry is in the context of the signal-to-noise ratio (SNR, and therefore rate) of the UE $\rightarrow$ RS link being different from that of the BS $\rightarrow$ RS link. 

The work in \cite{asym_ref_paulraj,asym_ref_popo} also considers a similar model with the additional assumptions that there are direct links between BS and UEs. Authors have shown that the rate performance can be improved by exploiting the direct links. 

Precoder design for the conventional symmetric non-regenerative TWR is an active area of research and is considered in \cite{twr_opt_pre_ref0,eurasip_af_mimo_ref,twr_opt_pre_ref1,twr_opt_pre_ref5,twr_opt_pre_ref2,twr_opt_pre_ref4,twr_opt_pre_ref3}. In \cite{twr_opt_pre_ref0}, the optimal beamforming precoder matrix is designed at the multi-antenna relay and the system capacity-region characterized for single-antenna source nodes. Precoders are designed in \cite{eurasip_af_mimo_ref} using the zero-forcing (ZF) and linear minimum mean square error (MMSE) criteria for a MIMO relay and MIMO source nodes. Optimal source and relay matrices are designed in \cite{twr_opt_pre_ref5} when all the nodes employ linear-MMSE receivers. A joint design of source and relay precoders is considered in \cite{twr_opt_pre_ref2} and \cite{twr_opt_pre_ref1} to minimize the MSE and maximize the sum-rate respectively. In \cite{twr_opt_pre_ref4}, a sub-optimal relay precoder to maximize the sum-rate is designed using the gradient-descent algorithm.  

\textbf{Contribution and organization}:  We now present the organization and key contributions of the paper.

1) A new transmission protocol is proposed to solve the problem of TWR with non-simultaneous downlink and uplink data traffic. The two-way asymmetric relay model is described in Section \ref{sys_model_sec_ref}. A non-regenerative relay is considered because of its operational simplicity \cite{owr_opt_pre_ref}. This kind of non-regenerative asymmetric TWR with MIMO nodes is being considered for the first time.

2) Designed a novel linear BI cancellation precoder at the relay; the precoder also triangulates the MAC- and BC-phase channel matrices. The precoder design is based on the singular-value-decomposition (SVD) and QR decomposition \cite{hornandjohnson} of MAC- and BC-phase channel matrices and is discussed in Section \ref{precoder_des_sec_ref}.

3) The WSR maximization problem for the proposed precoder is shown to be a geometric program in the high-SNR regime in Section \ref{sum_rate_max_sec_ref}. Though the idea of casting the sum-rate maximization as a geometric program has been used in context of point-to-point wireless systems  in \cite{gp_tut_ref2} and conventional one-way relay based systems in \cite{gp_owr_ref}, it is important to note that its application to the addressed scenario in the first. The present work is different from \cite{gp_owr_ref} as we study the WSR maximization instead of the sum-rate maximization. Also, the MAC- and BC-phase channel matrices in asymmetric TWR are coupled together, different from one-way relaying. This makes it relatively harder to show that the WSR maximization is indeed a convex optimization program. In addition, the framework developed for studying the WSR problem is also used to solve relay-power minimization under certain rate and SNR constraints at the BS and RUE. 

4) The performance gain of the proposed protocol is analysed using Monte Carlo simulations in Section \ref{result_sec_ref} in two steps: (a) Performance improvement achieved by the proposed precoder is demonstrated over the conventional ZF- and MMSE-based solutions. (b) Performance gain of asymmetric TWR with the proposed precoder is compared with the one-way relaying and single-hop (direct) transmission in a cellular framework. It is shown that the proposed protocol outperforms the other two techniques by significant margin. 

\textbf{Notation}: Bold upper- and lower-case letters are used to denote matrices and column vectors, respectively. For a matrix $\mathbf{A}$, Tr$\left(\mathbf{A} \right)$, $\mathbf{A}^T$ and $\mathbf{A}^H$ denote its trace, transposition and conjugate-transposition, respectively. $\mathbf{I}_n$ denotes an $n \times n$ identity matrix. \textsf{diag (}$x_1, \cdots, x_n$) denotes a diagonal matrix with $x_1, \cdots, x_n$ as the diagonal elements. $\|\mathbf {x}\|$ denotes the $l_2$ norm of a vector $\mathbf {x}$ and $\mathbf{x}^*$ denotes its complex conjugation. The notation $\mathbf{x} \sim \mathcal{CN}(\mathbf{0},\mathbf{\Sigma)}$ denotes that $\mathbf{x}$ is a circularly-symmetric complex Gaussian random vector with covariance matrix $\mathbf{\Sigma}$. $\mathbb{E}$(\textperiodcentered) is used to denote the expectation operator. $|$c$|$ denotes the magnitude of a complex scalar. $\log_2$($\cdot$) is denoted as $\log(\cdot$).
 
\section{System model and protocol description for Asymmetric Two-Way relaying}
\label{sys_model_sec_ref}
A communication model for asymmetric relaying is illustrated in Fig. \ref{2WayRelaying3Node}. Here we assume that there are two UEs, TUE and RUE, which communicate with the BS through a non-regenerative half-duplex relay. During MAC phase, BS and TUE simultaneously transmit to the relay. The relay transmits a linear function of the received signal to the BS and RUE during the BC phase. We assume that there are no direct links between the BS and the two UEs. Also, the BS and two UEs have M antennas each while the relay has $N \ge 2M$ antennas. We make an assumption frequently made in the literature that only the relay has complete instantaneous channel state information (CSI) during MAC and BC phases while other nodes have CSI during the BC phase alone \cite{relay_csi_sup_nw_code,eurasip_af_mimo_ref,twr_opt_pre_ref4}. 

Let $\mathbf{y}_r$ be the $N \times 1$ received signal at the relay during MAC phase. Let $\mathbf{x}_u$ and $\mathbf{x}_b$ denote the $M \times 1$ data-vectors transmitted by the TUE and BS respectively. Then,
\begin{align}
 \mathbf{y}_r &= \mathbf{H}_{u}\mathbf{x}_u+ \mathbf{H}_{b}\mathbf{x}_b +\mathbf{n}_r.
\label{relay_rx_sig}
 \end{align}
Here $\mathbf {H}_u \text{ and } \mathbf{H}_b \in \mathbb{C}^{N \times M}$ are the uplink channels observed by the relay from the TUE and BS, respectively. The data vectors $\mathbf{x}_u$ and $\mathbf{x}_b$ can be thought of as $M$ parallel data streams transmitted each by TUE and BS and are assumed to be distributed as $\mathcal{CN}(\mathbf{0},\mathbf{\Sigma}_u)$ and $\mathcal{CN}(\mathbf{0},\mathbf{\Sigma}_b)$, respectively. Here $\mathbf{\Sigma}_u = \frac{P_u}{M}\mathbf{I}_M = \rho_u\mathbf{I}_M$ and $\mathbf{\Sigma}_b = \frac{P_b}{M}\mathbf{I}_M=\rho_b\mathbf{I}_M$. Also, $P_u$ and $P_b$ denote the transmit power of the TUE and BS, respectively. The $\mathbf{n}_r \in \mathbb{C}^{N \times 1} $ is the noise vector at the relay and is assumed to be distributed as $\mathcal{CN}(\mathbf{0},\sigma_r^2\mathbf{I}_{N})$. For the ease of precoder design in the sequel, we express the signal received at the relay in \eqref{relay_rx_sig} in an equivalent matrix form.
\begin{align}
\mathbf{y}_r = \mathbf{Hx} + \mathbf{n}_r.
\label{comp_relay_sig_ref}
\end{align}
The matrix $\mathbf{H} = \left[\mathbf {H}_u \text{  } \mathbf {H}_b\right]$ is the composite uplink channel and the vector $\mathbf{x} = [\mathbf{x}_u^T \text{ } \mathbf{x}_b^T]^T$ with $\mathbb{E}(\mathbf{x}\mathbf{x}^H) = \mathbf{Q}$ = \textsf{diag}$(\mathbf{\Sigma}_u$, {$\mathbf{\Sigma}_b)$. During BC phase, the relay performs linear processing on the received signal by multiplying it with a precoder matrix $\mathbf{W} \in \mathbb{C}^{N \times N} $. The $N\times 1$ signal vector to be transmitted from the relay is therefore given as
\begin{align}
\mathbf{x}_r &= \mathbf{W}\mathbf{y}_r.
\label{relay_tx_signal}
\end{align}
The precoder matrix $\mathbf{W}$ is subjected to the average power constraint of the relay:
\begin{align}
P_r &\ge \text{Tr}\left(\mathbb{E}(\mathbf{x}_r\mathbf{x}_r^H)\right) \nonumber\\
&= \text{Tr}\left(\mathbf{WHQ}\mathbf{H}^H\mathbf{W}^H +\sigma_r^2\mathbf{W}\mathbf{W}^H\right). \label{relay_pow_const_eq_ref} 
\end{align}
The signals received by RUE and BS, $\mathbf{y}_u \text{ and }\mathbf{y}_b$, respectively, during BC phase are given as 
\begin{align}
\mathbf{y}_i &=\mathbf{G}_{i}\mathbf{x}_r + \mathbf{n}_i,\ \  i = u,b. \label{ue_rx_sig_init_eq_ref} 
\end{align}
The noise vectors $\mathbf{n}_i$ are $ \sim \mathcal{CN}(\mathbf{0},\sigma^2\mathbf{I}_{M})$. Here $\mathbf {G}_u \text{ and } \mathbf{G}_b \in \mathbb{C}^{M \times N}$ are the downlink channels observed by the RUE and BS, respectively. 
The signal received by the RUE and BS during the BC phase in \eqref{ue_rx_sig_init_eq_ref} are stacked to form a vector $\mathbf{y}$ such that
\begin{equation}
 \mathbf{y}= \mathbf{Gx}_r + \mathbf{n}.
 \label{bc_ch_mat_eq_ref}
\end{equation}
Here the vector $\mathbf{y} = {\left[\mathbf{y}_u^T \text{  } \mathbf{y}_b^T \right]}^T$ and $\mathbf{G} = {\left[\mathbf{G}_u^T \text{  } \mathbf{G}_b^T \right]}^T$ is the composite downlink channel matrix. Also, $\mathbf{n} = [\mathbf{n}_u^T \text{ } \mathbf{n}_b^T]^T$ $\sim \mathcal{CN}(\mathbf{0},\sigma^2\mathbf{I}_{2M})$. 
\section{Precoder design}
\label{precoder_des_sec_ref}
This section deals with the design of precoder which cancels the BI and triangulates the end-to-end channels observed by the RUI and BS. Towards this end, we first develop the structure of the precoder matrix $\mathbf{W}$, wherein it is decomposed into an uplink precoder matrix $\mathbf{F}$, permutation and power-distribution matrix $\mathbf{D}$, and a downlink precoder matrix $\mathbf{M}$ as: 
\begin{equation}
\mathbf{W} = \mathbf{M}\mathbf{D}\mathbf{F}. 
\label{pre_def_eq_ref}
\end{equation}
Here $\mathbf{M} \in \mathbb{C}^{N \times 2M}$ and $\mathbf{F} \in \mathbb{C}^{2M \times N}$ are the downlink and uplink precoders, respectively and are designed to completely cancel the BI for RUE. Precoders $\mathbf{M}$ and $\mathbf{F}$ are further decomposed into $\mathbf{M} = [\begin{array}{cc} \mathbf{M}_u &\mathbf{M}_b \end{array}]$ and $\mathbf{F} = {[\begin{array}{cc} \mathbf{F}_u^T &\mathbf{F}_b^T \end{array}]}^T $, respectively. Here $\mathbf{M}_u, \mathbf{M}_b \in \mathbb{C}^{N \times M}$ and $\mathbf{F}_u, \mathbf{F}_b \in \mathbb{C}^{M \times N}$ are termed as individual downlink and uplink precoders, respectively. The matrix $\mathbf{D}$ is defined as 
\begin{equation}
\mathbf{D} = \left[\begin{array}{ccrr} \mathbf{0} &\mathbf{D}_u \\\mathbf{D}_b &\mathbf{0} \end{array}\right]. 
\label{d_mat_def_ref}
\end{equation}
The constituent matrix $\mathbf{D}_u$ (resp. $\mathbf{D}_b$) is designed later to triangulate the end-to-end channels observed by the RUE (resp. BS). We will show that the channel triangularization will reduce WSR maximization problem to the power allocation by the relay to the RUE and BS. Therefore, matrix $\mathbf{D}_u$ (resp. $\mathbf{D}_b$) in addition, also determine the power distribution from the relay to the RUE (resp. BS). It is worth mentioning that the matrix $\mathbf{D}$ also permutes the receive signal at the relay. 

Before designing the individual precoder matrices, we summarize the design steps for the precoder $\mathbf{W}$:

1) Design $\mathbf{M}$ and $\mathbf{F}$ to cancel the BI observed by the RUE.

2) Design $\mathbf{D}_u$ and $\mathbf{D}_b$ to triangulate the end-to-end channels observed by the RUE and BS respectively, and maximize the WSR.
Henceforth, $\mathbf{M}$ and $\mathbf{F}$ will be referred as the downlink and uplink BI cancellation precoders, respectively, and $\mathbf{D}$ will be referred as the channel triangularization precoder.
\subsection{Back-propagating interference cancellation precoder design}
\label{lq_precoder_sec_ref}
To design the BI cancellation precoders, the vector $\mathbf{y}$ in \eqref{bc_ch_mat_eq_ref} can be re-expressed by substituting the expressions of $\mathbf{y}_r$, $\mathbf{x}_r$ and $\mathbf{W}$ from \eqref{comp_relay_sig_ref}, \eqref{relay_tx_signal} and \eqref{pre_def_eq_ref}, respectively.
\begin{align}
\mathbf{y} & = \mathbf{GW}\left(\mathbf{Hx} + \mathbf{n}_r\right)+ \mathbf{n}  \nonumber \\
& = \mathbf{GW}\mathbf{Hx} + \mathbf{GW}\mathbf{n}_r+ \mathbf{n}  \nonumber \\
& = \underbrace{\mathbf{GM}}_{\widetilde{\mathbf{G}}} \mathbf{D} \underbrace{\mathbf{FH}}_{\widetilde{\mathbf{H}}} \mathbf{x} + \underbrace{\mathbf{GW}\mathbf{n}_r+ \mathbf{n}}_{\tilde{\mathbf{n}}}  \nonumber \\
& = \widetilde{\mathbf{G}} \mathbf{D}\widetilde{\mathbf{H}}\mathbf{x}+ \tilde{\mathbf{n}}. \label{pre_design_eq_ref2}
\end{align}
In order that the signal received by RUE is interference-free, we state the following lemma.   
\begin{lemma}
Precoders $\mathbf{M}$ and $\mathbf{F}$ should be designed such that $\widetilde{\mathbf{G}} \in \mathbb{C}^{2M \times 2M}$ and $\widetilde{\mathbf{H}} \in \mathbb{C}^{2M \times 2M}$ are block lower- and upper-triangular matrices, respectively. 
\end{lemma}
\begin{IEEEproof}
With the block lower- and upper-triangular matrices $\widetilde{\mathbf{G}}$ and $\widetilde{\mathbf{H}}$, \eqref{pre_design_eq_ref2} will become:
\begin{align}
 \mathbf{y}
&  = \left[\begin{array}{ccrr} \widetilde{\mathbf{G}}_u & \mathbf{0} \\ \widetilde{\mathbf{G}}_n  &\widetilde{\mathbf{G}}_b \end{array}\right]
   \left[\begin{array}{ccrr} \mathbf{0} & \mathbf{D}_u \\\mathbf{D}_b  &\mathbf{0}  \end{array}\right]
   \left[\begin{array}{ccrr} \widetilde{\mathbf{H}}_b & \widetilde{\mathbf{H}}_n \\ \mathbf{0}  &\widetilde{\mathbf{H}}_u \end{array}\right]
    \left[\begin{array}{cr} \mathbf{x}_u \\ \mathbf{x}_b \end{array}\right]+ \tilde{\mathbf{n}} \nonumber, \\
& = \left[\begin{array}{cr}  (\widetilde{\mathbf{G}}_u \mathbf{D}_u \widetilde{\mathbf{H}}_u) \mathbf{x}_b\\ (\widetilde{\mathbf{G}}_b\mathbf{D}_b\widetilde{\mathbf{H}}_b)\mathbf{x}_u + (\widetilde{\mathbf{G}}_n\mathbf{D}_u\widetilde{\mathbf{H}}_u+\widetilde{\mathbf{G}}_b\mathbf{D}_b\widetilde{\mathbf{H}}_n)\mathbf{x}_b \end{array}\right] + \tilde{\mathbf{n}}. \label{equvi_ch_eq_ref}
\end{align}
Here $\widetilde{\mathbf{G}}_i,\widetilde{\mathbf{H}}_i \in \mathbb{C}^{M \times M}$ and $i \in \{u,b,n\}$. The vector $\mathbf{y} = {\left[\mathbf{y}_u^T \text{  } \mathbf{y}_b^T \right]}^T$. Recall that the TUE and BS transmitted $\mathbf{x}_u$ and $\mathbf{x}_b$ respectively during MAC phase.  It can be seen that RUE can detect its desired data $\mathbf{x}_b$ from its received signal $\mathbf{y}_u$ (first block-row in \eqref{equvi_ch_eq_ref}) without any interference. 
\end{IEEEproof}
The BS will as usual be able to cancel the self-interference $\mathbf{x}_b$ from its received signal $\mathbf{y}_b$ (second block-row in \eqref{equvi_ch_eq_ref}) and detect its desired data $\mathbf{x}_u$.\footnote{It is assumed that the BS has necessary channel knowledge to cancel the self-interference as commonly assumed in the TWR literature \cite{twr_opt_pre_ref1,twr_opt_pre_ref2,twr_opt_pre_ref5}.} 
\begin{remark}
RUE now needs to only estimate its own effective channel as its BI is completely cancelled. The CSI requirement at the RUE is thus considerably reduced.
\end{remark}
We next consider a technique to design the precoder matrices $\mathbf{F}$  and $\mathbf{M}$.
\iftoggle{DOUBLE_COL}{
    \begin{figure*}[!bhp]
\noindent\makebox[\linewidth]{\rule{\textwidth}{0.5pt}}    
    \begin{center}
    \includegraphics[width=6.5in]{asym_relay_mimo_tx_rx.eps} 
    \caption{\small TUE and BS transmitter chains, RUE and BS receiver chains for asymmetric TWR.}
    \label{tx_rx_chain_fig_ref}
    \end{center}
    \end{figure*}
}{}

\textbf{Design of precoder matrices $\mathbf{F}$  and $\mathbf{M}$}: To design $\mathbf{F}$ and $\mathbf{M}$, matrices $\widetilde{\mathbf{H}}$ and $\widetilde{\mathbf{G}}$ in \eqref{pre_design_eq_ref2} are re-expressed by plugging the expressions of $\mathbf{H}$, $\mathbf{G}$ and $\mathbf{M}$, $\mathbf{F}$ from \eqref{comp_relay_sig_ref}, \eqref{bc_ch_mat_eq_ref} and \eqref{pre_def_eq_ref}, respectively. 
\begin{align}
\widetilde{\mathbf{H}} = \left[\begin{array}{ccrr} \mathbf{F}_u\mathbf{H}_u & \mathbf{F}_u\mathbf{H}_b \\ 
\mathbf{F}_b\mathbf{H}_u  &\mathbf{F}_b\mathbf{H}_b \end{array}\right],\text{ } 
\widetilde{\mathbf{G}} = \left[\begin{array}{ccrr} \mathbf{G}_u\mathbf{M}_u & \mathbf{G}_u\mathbf{M}_b \\ 
\mathbf{G}_b\mathbf{M}_u  &\mathbf{G}_b\mathbf{M}_b \end{array}\right]. \label{hg_tilde_eq_ref}
\end{align}
In order that the matrix $\widetilde{\mathbf{H}}$ is block upper-triangular, the precoder matrix $\mathbf{F}_b$ be designed such that $\mathbf{F}_b\mathbf{H}_u=\mathbf{0}$. This implies that $\mathbf{F}_b$ should belong to the left null-space of ${\mathbf{H}}_u$.\footnote{The left null-space of a matrix $ \mathbf{H}$ contains vectors $\mathbf{v}$ such that $\mathbf{v}^H\mathbf{H}=\underline{\mathbf{0}}$.} To this end, we define the SVD of $\mathbf{H}_u$ as
\begin{equation}
\mathbf{H}_u  = \left[\begin{array}{cc}\mathbf{U}^{(1)}_{{\mathbf{H}}_u} &\mathbf{U}^{(0)}_{{\mathbf{H}}_u} \end{array}\right]
			\begin{array}{c} \sum_{{\mathbf{H}}_u} \end{array} \mathbf{V}^{H}_{{\mathbf{H}}_u}, 
\end{equation}
where $\mathbf{U}^{(1)}_{{\mathbf{H}}_u} \in \mathbb{C}^{N \times M}$ contains the first $M$ left singular vectors and $\mathbf{U}^{(0)}_{{\mathbf{H}}_u} \in \mathbb{C}^{N \times \bar{N}}$ contains the last $\bar{N} = N - M$ left singular vectors. Note that $N \ge 2M$. It is known that  the columns of $\mathbf{U}^{(0)}_{{\mathbf{H}}_u}$ form an orthonormal basis set for the left null-space of ${\mathbf{H}}_u$ \cite{hornandjohnson}. We therefore choose $\mathbf{F}_b$ as the first $M$ columns of $\mathbf{U}^{(0)}_{{\mathbf{H}}_u}$ i.e.,  $\mathbf{F}_b = \mathbf{U}^{(0)H}_{{\mathbf{H}}_u}\left( m\right), m= 1, \cdots, M$. Precoder $\mathbf{F}_u$ can  be chosen as any arbitrary matrix which does not affect the block upper-triangular structure of the matrix $\widetilde{\mathbf{H}}$. Without loss of generality (w.l.o.g) we choose $\mathbf{F}_u = \mathbf{U}^{(1)H}_{{\mathbf{H}}_u}$. The uplink precoder $\mathbf{F}$ is therefore given as 
\begin{equation}
\mathbf{F} = \left[\begin{array}{rcc} & {\mathbf{U}^{(1)}_{{\mathbf{H}}_u}}^*  &{\mathbf{U}^{(0)}_{{\mathbf{H}}_u}}^*(m) \end{array}\right]^T, \ \ m = 1,\cdots, M.
\label{ul_bpi_can_pre_ref}
\end{equation}
We next design the downlink BI cancellation precoder $\mathbf{M}$. For the matrix $\widetilde{\mathbf{G}}$ to be block lower-triangular, it can be seen from \eqref{hg_tilde_eq_ref} that $\mathbf{M}_b$ should be in the null-space of ${\mathbf{G}}_u$ i.e., ${\mathbf{G}}_u\mathbf{M}_b=\mathbf{0}$. The SVD of ${\mathbf{G}}_u$ is performed to determine its null-space.
\begin{equation}
{\mathbf{G}}_u  = \mathbf{U}_{{\mathbf{G}}_u} \begin{array}{c} \sum_{{\mathbf{G}}_u} \end{array} 
			    \left[\begin{array}{cc}\mathbf{V}^{(1)}_{{\mathbf{G}}_u} &\mathbf{V}^{(0)}_{{\mathbf{G}}_u} \end{array}\right]^{H},
\end{equation}
where $\mathbf{V}^{(1)}_{{\mathbf{G}}_u} \in \mathbb{C}^{N \times M}$ contains the first $M$ right singular vectors and $\mathbf{V}^{(0)}_{{\mathbf{G}}_u} \in \mathbb{C}^{N \times \bar{N}}$ contains the last $\bar{N} = N - M$ right singular vectors. The columns of  $\mathbf{V}^{(0)}_{{\mathbf{G}_u}}$ form an orthonormal basis set for the null-space of ${\mathbf{G}_u}$ \cite{hornandjohnson}. We therefore choose first M columns of $\mathbf{V}^{(0)}_{{\mathbf{G}_u}}$ for the precoder matrix $\mathbf{M}_b$. It is clear from \eqref{hg_tilde_eq_ref} that the precoder matrix $\mathbf{M}_b$ can  be chosen as any arbitrary matrix which does not affect the block lower-triangular structure of the matrix $\widetilde{\mathbf{G}}$. The downlink precoder $\mathbf{M}_u$ is therefore chosen w.l.o.g. as $\mathbf{V}^{(1)}_{{\mathbf{G}_u}}$.\footnote{We later show in Section \ref{sum_rate_max_sec_ref} that the unitary structure of $\mathbf{M}$ and $\mathbf{F}$ matrices is desired in casting the WSR maximization as a convex optimization program.} The downlink precoder $\mathbf{M}$ can thus be written as 
\begin{equation}
\mathbf{M} = \left[\begin{array}{rcc} & {\mathbf{V}^{(1)}_{{\mathbf{G}}_u}}  &{\mathbf{V}^{(0)}_{{\mathbf{G}}_u}}(m) \end{array}\right], \ \ m=1,\cdots, M.
\label{dl_bpi_can_pre_ref}
\end{equation}
\subsection{Channel Triangularization precoder design}
This section deals with the design of channel triangularization precoder $\mathbf{D}$. The structure of the channel triangularization precoder is such that the $M$ parallel streams are decoupled at the respective receivers with minimal signal processing. This is critical for the RUE which has limited processing capabilities. The proposed precoder structure also reduces the WSR maximization to power allocation problem at the relay, which can be cast as a convex optimization problem in the high SNR regime.

To design $\mathbf{D}$, we note from \eqref{equvi_ch_eq_ref} that the signal received by the RUE is

\begin{align}
\widehat{\mathbf{y}}_u = \mathbf{y}_u = \left(\widetilde{\mathbf{G}}_u \mathbf{D}_u \widetilde{\mathbf{H}}_u\right) \mathbf{x}_b + \tilde{\mathbf{n}}_u.
\label{iue_rx_sig}
\end{align}
Similarly, signal observed by the BS after cancelling the self- interference is
\begin{align}
\widehat{\mathbf{y}}_b = \left(\widetilde{\mathbf{G}}_b \mathbf{D}_b \widetilde{\mathbf{H}}_b\right)\mathbf{x}_u + \tilde{\mathbf{n}}_b.
\label{bs_rx_sig}
\end{align}
The vectors $\tilde{\mathbf{n}}_u \sim \mathcal{CN}(\mathbf{0},\mathbf{\Sigma_{\tilde{n}_u}})$ and $\tilde{\mathbf{n}}_b\sim \mathcal{CN}(\mathbf{0},\mathbf{\Sigma_{\tilde{n}_b}})$ are the effective noise observed by the RUE and BS with the covariance matrices given respectively as 
\begin{align}
 \mathbf{\Sigma_{\tilde{n}_u}} &= \sigma_r^2\left\{\widetilde{\mathbf{G}}_u\mathbf{D}_u(\widetilde{\mathbf{G}}_u\mathbf{D}_u)^H\right\} + \sigma^2\mathbf{I}_M, \label{bs_ue_noise_cov_mat}\\
 \mathbf{\Sigma_{\tilde{n}_b}} &= \sigma_r^2\left\{\widetilde{\mathbf{G}}_n\mathbf{D}_u(\widetilde{\mathbf{G}}_n\mathbf{D}_u)^H+
						\widetilde{\mathbf{G}}_b\mathbf{D}_b(\widetilde{\mathbf{G}}_b\mathbf{D}_b)^H\right\} + \sigma^2\mathbf{I}_M.
						\nonumber
\end{align}
The above matrices are calculated from \eqref{equvi_ch_eq_ref} by using the fact that the uplink BI cancellation precoder $\mathbf{F}$ has orthonormal rows by design.

It can be seen from \eqref{iue_rx_sig} and \eqref{bs_rx_sig} that the signal received by the RUE and BS is a function of the precoders $\mathbf{D}_u$ and $\mathbf{D}_b$, respectively. This leads to considerable simplification in the channel triangularization precoder design as $\mathbf{D}_u$ and $\mathbf{D}_b$ can be designed to triangulate the channel for RUE and BS separately. We next define the structure of precoders $\mathbf{D}_u$ and $\mathbf{D}_b$ in the following equation. 
\begin{align}
 \mathbf{D}_i = \mathbf{\Pi}_i\mathbf{\Delta}_i\mathbf{\Theta}_i.
 \label{isi_pre_eq_ref}
\end{align}
Here $i\in\{u,b\}$. The matrix $\mathbf{\Delta}_i \in \mathbb{R}^{M \times M}$ is an anti-diagonal matrix with non-negative variables $\sqrt{\delta}_{i,m}$, $m=1,\cdots, M$ as its elements. These variables decide power distribution across $M$ streams and are optimized later to maximize the WSR for the system. The matrices $\{\mathbf{\Pi}_i \text { and } \mathbf{\Theta}_i\} \in \mathbb{C}^{M \times M}$ are designed to triangulate the BC- and MAC-phase channels, respectively.
\iftoggle{DOUBLE_COL}{
\begin{figure*}[!b]
\noindent\makebox[\linewidth]{\rule{\textwidth}{0.5pt}}
\begin{eqnarray}
\begin{aligned}
\text{\text{SNR}}_{b,\widetilde m} &= \frac{\delta_{b,m}\Big|{[\mathbf{L}_b]}_{m,m}{[\mathbf{R}_b]}_{\widetilde m,\widetilde m}\Big|^2\rho_u}
{\sigma_r^2\displaystyle\sum_{k=1}^M\left\{\delta_{u,k}\left({[\widetilde{\mathbf{G}}_n\mathbf{\Pi}_u]}_{m,k}{[{\widetilde{\mathbf{G}}_n\mathbf{\Pi}_u}]}_{{m,k}}^*\right) + 
\delta_{b,k}\left({[\mathbf{L}_b]}_{m,k}{[\mathbf{L}_b]}_{m,k}^*\right)\right\} + \sigma^2},\\
\text{\text{SNR}}_{u,\widetilde m}&= \frac{\delta_{u,m}\Big|{[\mathbf{L}_u]}_{m,m}{[\mathbf{R}_u]}_{\widetilde m,\widetilde m}\Big|^2\rho_b}
{\sigma_r^2 \displaystyle\sum_{k=1}^M \delta_{u,k}\Big({[\mathbf{L}_u]}_{m,k}{[\mathbf{L}_u]}_{m,k}^*\Big) + \sigma^2}.\label{bs_ue_fin_snr_sic}
\end{aligned}
\end{eqnarray}
\end{figure*}
}{}
The signal received by the RUE and BS can be re-expressed by plugging the expressions of $\mathbf{D}_u$ and $\mathbf{D}_b$ from \eqref{isi_pre_eq_ref} as follows 
\begin{align}
 \widehat{\mathbf{y}}_i &= \underbrace{\widetilde{\mathbf{G}}_i\mathbf{\Pi}_i\mathbf{\Delta}_i\mathbf{\Theta}_i\widetilde{\mathbf{H}}_i}_{\mathbf{C}_i}\mathbf{x}_{\bar{i}} + 
 \tilde{\mathbf{n}}_i, \\
 & = \mathbf{C}_i\mathbf{x}_{\bar{i}} + \tilde{\mathbf{n}}_i.  
\end{align}
If $\mathbf{\Pi}_i$ and $\mathbf{\Theta}_i$ are designed such that $\widetilde{\mathbf{G}}_i\mathbf{\Pi}_i$ and $\mathbf{\Theta}_i\widetilde{\mathbf{H}}_i$ are \textit{lower-triangular} and \textit{upper-triangular} respectively,\footnote{To avoid stating repeatedly, we assume that $i\in\{u,b\}$ for the rest of discussions in the sequel. Also $\bar{i} = u$ for $i = b$ and $\bar{i} = b$ for $i = u$.}, the end-to-end channel observed by $\mathbf{x}_{\bar{i}}$ (i.e., $\mathbf{C}_i$) will have a reflected-lower-triangular structure as shown below:
\begin{equation}
\left[\begin{array} {crrrrr} \widehat{y}_{i,1} \\\widehat{y}_{i,2} \\\vdots\\\widehat{y}_{i,M-1} \\\widehat{y}_{i,M}\end{array}\right] = 
\left[\begin{array}{cccccrrrrr} 
0 &0 &\cdots &0 &\times\\
0 & & &\times &\times\\
\vdots & &\iddots &\vdots &\vdots\\
0 &\times &\cdots &\times &\times\\
\times &\times &\cdots &\times &\times
\end{array}\right]
\left[\begin{array} {crrrrr} {x}_{\bar{i},1} \\{x}_{\bar{i},2} \\\vdots\\{x}_{\bar{i},M-1} \\{x}_{\bar{i},M}\end{array}\right] + \tilde{\mathbf{n}}_i. 
\label{rx_sig_str_eq_ref}
\end{equation}
With this received signal structure, $(M-k)$th stream is detected by subtracting the interference from $(M-k+1)$th to $M$th streams, in a manner similar to successive interference cancellation (SIC) \cite{vblast_paper}. Here $k = 1,\cdots,M-1$. Note that the last (i.e., $M$th) stream does not observe any interference and is detected first. It is important to note that the anti-diagonal structure of power allocation matrix $\mathbf{\Delta}_i$ plays a crucial role in reducing $\widehat{\mathbf{y}}_i$ to the above form. The complete receiver processing for the BS and RUE is shown in the transceiver chains in Fig. \ref{tx_rx_chain_fig_ref}. The BS receiver first performs BI cancellation followed by the SIC to decode its $M$ streams. Since the proposed precoder completely cancels the BI observed by the RUE, BI cancellation block is replaced by a pass-through $\mathbf{I}_M$ block in the RUE receiver. RUE thus performs only SIC to decode its $M$ streams. 

\iftoggle{SINGLE_COL}{
    \begin{figure}
    \begin{center}
  \includegraphics[width=6.4in]{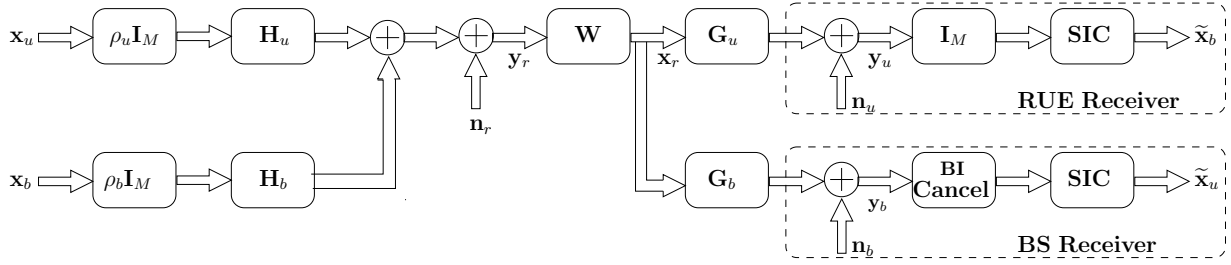} 
  \caption{\small RUE and BS transceiver chains for asymmetric TWR.}
    \label{tx_rx_chain_fig_ref}
    \end{center}
    \end{figure}
}{}

\textbf{Design of $\mathbf{\Pi}_i$ and $\mathbf{\Theta}_i$}:
Recall that $\mathbf{\Pi}_i$ should be designed such that $\widetilde{\mathbf{G}}_i\mathbf{\Pi}_i$ has lower-triangular structure. To design $\mathbf{\Pi}_i$, the matrix $\widetilde{\mathbf{G}}_i$ is decomposed into a lower-triangular matrix and a unitary matrix using the LQ decomposition \cite{hornandjohnson}. The LQ decomposition of $\widetilde{\mathbf{G}}_i$ is denoted as 
 \begin{equation}
 \widetilde{\mathbf{G}}_i = \mathbf{L}_i\widehat{\mathbf{Q}}_{i},
\label{lq_eq_ref1}
\end{equation}
 where $\mathbf{L}_i \in \mathbb{C}^{M\times M}$ is a lower-triangular matrix and $\widehat{\mathbf{Q}}_{i} \in \mathbb{C}^{M \times M}$ is a unitary matrix.  For $\widetilde{\mathbf{G}}_i\mathbf{\Pi}_i$ to be lower-triangular, choose $\mathbf{\Pi}_i =\widehat{\mathbf{Q}}^H_{i}$. Similarly, $\mathbf{\Theta}_i$ should be designed such that $\mathbf{\Theta}_i\widetilde{\mathbf{H}}_i$ has an upper-triangular structure. To design $\mathbf{\Theta}_i$, $\widetilde{\mathbf{H}}_i$is  decomposed into a unitary matrix and an upper-triangular matrix using QR decomposition \cite{hornandjohnson}. We denote the QR decomposition of $\widetilde{\mathbf{H}}_i$ as 
\begin{equation}
 \widetilde{\mathbf{H}}_i =  \mathbf{Q}_{i}\mathbf{R}_i,
\end{equation}
where $\mathbf{Q}_{i} \in \mathbb{C}^{M \times M}$ is a unitary matrix and $\mathbf{R}_i \in \mathbb{C}^{M \times M}$ is an upper-triangular matrix. To reduce $\mathbf{\Theta}_i\widetilde{\mathbf{H}}_i$ to an upper-triangular matrix, we choose $\mathbf{\Theta}_i = \mathbf{Q}^H_{i}$.  
 The precoder $\mathbf{D}_i$ is therefore given as
 \begin{equation}
 \mathbf{D}_i = \widehat{\mathbf{Q}}^{H}_{i} \mathbf{\Delta}_i \mathbf{Q}^{H}_{i} =  
 \left(\mathbf{Q}_{i} \mathbf{\Delta}^T \widehat{\mathbf{Q}}_{i}\right)^H.
\label{cp_pre_ref}
 \end{equation}
SNRs observed by $\widetilde m$th stream of BS and RUE can be calculated by using \eqref{iue_rx_sig}, \eqref{bs_rx_sig}, \eqref{bs_ue_noise_cov_mat} and are given respectively as
\begin{equation}
\begin{aligned}
\text{\text{SNR}}_{b,\widetilde m} &= \frac{\delta_{b,m}\Big|{[\mathbf{L}_b]}_{m,m}{[\mathbf{R}_b]}_{\widetilde m,\widetilde m}\Big|^2\rho_u}
{\sigma_r^2 \left(\left[\mathbf{T}_n\mathbf{T}_n^H\right]_{m,m}+ \left[\mathbf{T}_b\mathbf{T}_b^H\right]_{m,m}\right) + \sigma^2},\\
\text{\text{SNR}}_{u,\widetilde m} &= \frac{\delta_{u,m}\Big|{[\mathbf{L}_u]}_{m,m}{[\mathbf{R}_u]}_{\widetilde m,\widetilde m}\Big|^2\rho_b}{\sigma_r^2\left[\mathbf{T}_u\mathbf{T}_u^H\right]_{m,m} + \sigma^2}.
\label{bs_ue_ini_snr_sic}
\end{aligned}
\end{equation}
Here $\widetilde m= M - m + 1$ and $m=1,\cdots, M$.  Also, $\mathbf{T}_u = \widetilde{\mathbf{G}}_u\mathbf{D}_u$, $\mathbf{T}_n = \widetilde{\mathbf{G}}_n\mathbf{D}_u$ and $\mathbf{T}_b = \widetilde{\mathbf{G}}_b\mathbf{D}_b$. As both $\mathbf{\Theta}_u$ and $\mathbf{\Theta}_b$ are unitary matrices, SNR expressions can be further simplified and are given in \eqref{bs_ue_fin_snr_sic} \iftoggle{DOUBLE_COL}{(at the bottom of the page)}{}.   
\iftoggle{SINGLE_COL}{
\begin{eqnarray}
\begin{aligned}
\text{\text{SNR}}_{b,\widetilde m} &= \frac{\delta_{b,m}\Big|{[\mathbf{L}_b]}_{m,m}{[\mathbf{R}_b]}_{\widetilde m,\widetilde m}\Big|^2\rho_u}
{\sigma_r^2\displaystyle\sum_{k=1}^M\left\{\delta_{u,k}\left({[\widetilde{\mathbf{G}}_n\mathbf{\Pi}_u]}_{m,k}{[{\widetilde{\mathbf{G}}_n\mathbf{\Pi}_u}]}_{{m,k}}^*\right) + 
\delta_{b,k}\left({[\mathbf{L}_b]}_{m,k}{[\mathbf{L}_b]}_{m,k}^*\right)\right\} + \sigma^2},\\
\text{\text{SNR}}_{u,\widetilde m}&= \frac{\delta_{u,m}\Big|{[\mathbf{L}_u]}_{m,m}{[\mathbf{R}_u]}_{\widetilde m,\widetilde m}\Big|^2\rho_b}
{\sigma_r^2 \displaystyle\sum_{k=1}^M \delta_{u,k}\Big({[\mathbf{L}_u]}_{m,k}{[\mathbf{L}_u]}_{m,k}^*\Big) + \sigma^2}.\label{bs_ue_fin_snr_sic}
\end{aligned}
\end{eqnarray}
}{}
Note that the coefficients of power-distribution variables, $\delta_{u,m} \text{ and } \delta_{b,m} $, are non-negative, $\forall m$. This is possible because $\mathbf{\Theta}_u$, $\mathbf{\Theta}_b$ and uplink BI cancellation precoder $\mathbf{F}$ (cf. \eqref{ul_bpi_can_pre_ref})  are unitary matrices. This fact will be useful in proving the convexity of WSR optimization problem in the next section.
\begin{remark}
\textit{Channel parallelization}:
Instead of the channel triangularization approach discussed above, $\mathbf{D}_u$ and $\mathbf{D}_b$ can also be designed to perform the channel parallelization at the relay as follow: 
\begin{align}
\mathbf{D}_u = \widetilde{\mathbf{G}}_u^{-1}\mathbf{\Delta}_u\widetilde{\mathbf{H}}_u^{-1},\text{  } 
\mathbf{D}_b = \widetilde{\mathbf{G}}_b^{-1}\mathbf{\Delta}_b\widetilde{\mathbf{H}}_b^{-1}.
\label{zf_prec_ref}
\end{align}
This block-ZF approach will lead to simpler receiver architecture when compared to the channel triangularization approach, as there is no need to perform SIC. 
\end{remark}
\begin{remark}
\label{mu_ext_rem_ref}
\textit{Extension to multiple user-pair scenario}: The $M$ downlink data streams transmitted by the BS can be targeted to $M$ single-antenna users, RUE$_1\cdots$RUE$_M$. With the received signal structure in \eqref{rx_sig_str_eq_ref}, zero-forcing dirty-paper (ZF-DP) coding \cite{caire_bc_cap_ref} can be applied at the BS to ensure an interference-free channel for each of the $M$ RUEs. SNR observed by the $m$th RUE in the multiple user-pair scenario will be same as the SNR of $m$th stream in the single user-pair case (cf. \eqref{bs_ue_fin_snr_sic}) discussed before.  Similarly, $M$ independent uplink data streams transmitted by the TUE can be thought of as $M$ independent streams from $M$ single-antenna users, TUE$_1\cdots$TUE$_M$, each transmitting a single stream. BS will decode all the $M$ streams as usual with each stream observing the same SNR as in the single user-pair scenario. By applying ZF-DP coding at the BS, the proposed precoder can thus enable asymmetric two-way relay communication between a BS, $M$ single-antenna TUEs and $M$ single-antenna RUEs. Note that for single user-pair, ZF-DP is not required as RUE can decode all its $M$ streams by employing SIC.
\end{remark}

\section{Weighted Sum Rate maximization}
\label{sum_rate_max_sec_ref}
The WSR of the system is defined as 
\begin{align}
 R_\textsf{sum}(\boldsymbol{\delta}) &= \frac{1}{2}\sum_{i\in \{u,b\}}\sum_{m=1}^M w_{i,m}\log\left(1 + \text{SNR}_{i,m}(\boldsymbol{\delta})\right).\label{wsr_def_eq_ref} 
\end{align}
Here $\boldsymbol{\delta} \in \mathbb{R}^{2M\times 1}$ is a vector formed by stacking the power allocation variables i.e, $\boldsymbol{\delta} = [\delta_{u,1},\cdots, \delta_{u,M}, \delta_{b,1},\cdots, \delta_{b,M}]$. Here $w_{u,m}$ and $w_{b,m}$ are fixed non-negative scalar weights that allows QoS tradeoff for each uplink and downlink data streams. The factor of $1/2$ is due to the half-duplex constraint. In this section, we calculate $\delta_{u,m} $ and $\delta_{b,m} $ so as to maximize the WSR for the precoder design discussed above. The WSR maximization problem can be stated as 
\begin{equation}
\begin{aligned}
 & \underset{\boldsymbol{\delta}:\boldsymbol{\delta} \succeq 0} {\text{ Max. }} 
&& R_\textsf{sum}(\boldsymbol{\delta})\\
&\text{ s.t. } 
&& \eqref{relay_pow_const_eq_ref}
\label{af_opt_prob_def}
\end{aligned}
\end{equation}
The constraint in the optimization problem is imposed on the total transmit power of the relay as in \eqref{relay_pow_const_eq_ref}. Also, $\boldsymbol{\delta} \succeq 0$ implies that $\delta_{u,m}  \ge 0$ and $\delta_{b,m}  \ge 0,\ m = 1,\cdots,M$. The optimization problem in the present form is shown as non-convex in Appendix \ref{non_con_app_ref}. We next use the high-SNR approximation to cast the optimization problem as a geometric program (GP). A GP can be transformed into a convex program after a logarithmic change of variables. The objective function in \eqref{af_opt_prob_def} can be approximated at high SNR as  
\begin{align}
&\simeq \frac{1}{2}\sum_{m=1}^M \Big({w_{u,m}}\log\big(\text{SNR}_{u,m}(\boldsymbol{\delta})\big) + 
	 {w_{b,m}}\log\big(\text{SNR}_{b,m}(\boldsymbol{\delta})\big)\Big)\nonumber\\ 
&= \frac{1}{2}\log\Big(\prod_{m=1}^M({\text{SNR}_{u,m}(\boldsymbol{\delta})})^{w_{u,m}}({\text{\text{SNR}}_{b,m}(\boldsymbol{\delta})})^{w_{b,m}}\Big).
\label{af_sum_rate_relax_ref}
\end{align}
Maximizing the weighted sum-rate is thus equivalent to maximizing the product of SNRs or minimizing the product of inverse SNRs (denoted as ISNRs). Weighted sum-rate maximization problem is equivalent to
\begin{equation}
\begin{aligned}
 & \underset{\boldsymbol{\delta}\succeq 0} {\text{ Min. }} 
&&\prod_{m=1}^M ({\text{ISNR}_{u,m}(\boldsymbol{\delta})})^{w_{u,m}}({\text{ISNR}_{b,m}(\boldsymbol{\delta})})^{w_{b,m}}\\
&\text{ s.t. } 
&& \eqref{relay_pow_const_eq_ref}.
\label{final_af_opt_prob_ref}
\end{aligned}
\end{equation}
Here we have dropped the $1/2(\log)$ term from the objective function as $\log(\cdot)$ is a monotonically increasing function.
Before showing that the above optimization program can be formulated as a GP, we briefly explain the GP terminology from \cite{boyd_book} for the sake of completeness. We begin with a few definitions. A \textit{monomial} is a function $f : \mathbf{R}_{++}^n:\rightarrow \mathbf{R}$ of the form
\begin{equation}
 f(\mathbf{x}) = cx_1^{a_1}x_2^{a_2}\cdots x_n^{a_n},
\end{equation}
where $c > 0 \text{ and } a_j \in \mathbf {R}$. A sum of monomial functions is called a \textit{posynomial} function i.e.,
\begin{equation}
 f(\mathbf{x}) = \sum_{k=1}^K c_kx_1^{a_{1k}}x_2^{a_{2k}}\cdots x_n^{a_{nk}},
\end{equation}
where $c_k > 0$. Here $\mathbf{R}_{++}^n$ denotes the set of $n$-dimensional positive real  vectors. In a GP, the objective function and inequality constraints are posynomials and equality constraints are monomials. If $f_i:\mathbf{R}^n:\rightarrow \mathbf{R}$, $i=1,\cdots k$ are posynomial in $\mathbf{x}$ and $\phi:\mathbf{R}^k:\rightarrow \mathbf{R}$ is a posynomial with non-negative fractional exponents, then the composition $h(x) = \phi(f_1(\mathbf{x}),\cdots,f_k(\mathbf{x}))$ is defined as a generalized posynomial. In a generalized geometric program (GGP), the objective function and inequality constraints are generalized posynomials and equality constraints are monomials.

From the SNR expressions in \eqref{bs_ue_fin_snr_sic}, it can be easily seen that the \text{ISNR} is a valid posynomial function and the objective function therefore is a generalized posynomial. In order to show that the optimization problem can be solved as a GP, we first show that the power-constraint is a posynomial. 
This can be shown by proving the following lemma. 
\begin{lemma}
\label{pow_const_lemma_ref_sic}
Power constraint is a posynomial in $\delta_{u,m} $ and $\delta_{b,m} $, $m=1,\cdots, M$, if: 1) matrix $\mathbf{M}$ has orthonormal columns and the matrix $\mathbf{F}$ has orthonormal rows; and 2) matrices $\mathbf{\Pi}_i$ and $\mathbf{\Theta}_i$ are unitary. Here $i\in \{u,b\}$.
\end{lemma}
\begin{IEEEproof}
Refer to Appendix \ref{lq_qr_app_ref}.
\end{IEEEproof}
We next show that the generalized posynomial in the objective function can be handled in geometric programming framework by stating the following lemma. 
\begin{lemma}
\label{gen_posy_lemma_ref}
A generalized posynomial in the objective function can be expressed as equivalent posynomial constraints \cite{boyd_book}.
\end{lemma}
\begin{IEEEproof}
Refer to Appendix \ref{gen_posy_app_ref}.
\end{IEEEproof}
The optimization problem in \eqref{final_af_opt_prob_ref} can now be cast as a GP as both objective function and constraint are shown as posynomials; and can be solved using available software packages \cite{cvx_ref}. The high-SNR approximation is made in the literature and is applicable in scenarios where SNR is much larger than 0 dB \cite{gp_tut_ref2}. At low to medium SNRs, the approximation of $\log(1 + \text{SNR})$ as $\log({\text{SNR}})$ does not apply. Unlike \text{ISNR}, which is a posynomial, 1/(1+\text{SNR}) is not a posynomial. It is a ratio of two posynomials. One approach to handle a ratio of posynomials is the single condensation technique described in \cite{gp_tut_ref2}, where the posynomial in the denominator of the ratio is condensed to a monomial. Ratio of a posynomial and monomial is also a posynomial. The problem is then solved iteratively to improve the approximation at each step.  We use this approach to solve the optimization problem at low and moderate SNRs. 
\begin{remark}
With the knowledge that the ISNR$_u$, ISNR$_b$ and relay transmit power ($P_r$) are posynomials in $ \boldsymbol{\delta}$ for the designed precoder, we study another problem of practical interest as stated below. 
\begin{equation}
\begin{aligned}
 & \underset{\boldsymbol{\delta}\succeq 0} {\text{ Min. }} 
 P_r = f(\boldsymbol{\delta})\\
\text{ s.t. } 
& \sum_{m=1}^M\log({\text{SNR}_{u,m}(\boldsymbol{\delta})}) \ge r_u,\ \ 
\sum_{m=1}^M\log({\text{SNR}_{b,m}(\boldsymbol{\delta})}) \ge r_b.
\label{relay_pow_rate_opt_ref}
\end{aligned}
\end{equation}
The objective is to minimize the relay transmit power. The constraints specify QoS requirements in terms of data rates required by the TUE and RUE i.e., $r_b$ and $r_u$, respectively. The optimization problem in the above form is non-convex, but can be cast as a convex program by re-stating the constraints as follows: 
\begin{equation}
\begin{aligned}
 & \underset{\boldsymbol{\delta}\succeq 0} {\text{ Min. }} 
&& P_r = f(\boldsymbol{\delta})\\
&\text{ s.t. } 
&& \prod_{m=1}^M{\text{ISNR}_{u,m}(\boldsymbol{\delta})} \le 2^{-r_u},\ \
\prod_{m=1}^M{\text{ISNR}_{b,m}(\boldsymbol{\delta})} \le 2^{-r_b}.
\label{}
\end{aligned}
\end{equation}
\end{remark}

\begin{remark}
The QoS constraints in the optimization problem in \eqref{relay_pow_rate_opt_ref} can also be specified directly in terms of receive SNR required at the RUE and BS for each of their respective $M$ streams i.e., ${\text{SNR}_{u,m}(\boldsymbol{\delta})} \ge s_{u,m}$ and ${\text{SNR}_{b,m}(\boldsymbol{\delta})} \ge s_{b,m}, m = 1,\cdots,M$. The optimization problem with SNR QoS constraints is cast as
\begin{equation}
\begin{aligned}
 & \underset{\boldsymbol{\delta}\succeq 0} {\text{ Min. }} 
&& P_r = f(\boldsymbol{\delta})\\
&\text{ s.t. } 
&& {\text{ISNR}_{u,m}(\boldsymbol{\delta})} \le 1/s_{u,m}\text{ ,}\ \ 
{\text{ISNR}_{b,m}(\boldsymbol{\delta})} \le 1/s_{b,m}.
\label{relay_pow_snr_opt_ref}
\end{aligned}
\end{equation}
Note that the above optimization problem in \eqref{relay_pow_snr_opt_ref} is convex in any SNR regime due to convexity of the objective function and constraints at all SNRs, different from the other two problems in \eqref{final_af_opt_prob_ref} and \eqref{relay_pow_rate_opt_ref}.
\end{remark}

\section{Numerical results}
\label{result_sec_ref}
In this section, average WSR of the precoders is analysed using Monte Carlo simulations. We assume that the elements of uplink and downlink channels, $\mathbf {H}_i \text{ and } \mathbf{G}_i$, are independent and are distributed as $\mathcal{CN}(0,h_i^2)$ and $\mathcal{CN}(0,g_i^2)$ respectively, where $i \in \{u,b\}$.  We also assume that the nodes employ Gaussian signalling. The average WSR is obtained by solving the optimization problem in \eqref{af_opt_prob_def} and by averaging the WSR over $10^4$ statistically independent channel fading realizations.  The average WSR so obtained can be nearly achieved by employing capacity approaching error correcting codes and aggressive adaptive modulation as is done in the current cellular systems \cite{baker_book}, and hence can be considered reasonable. 
\subsection{WSR comparison of different precoders}
We first show the average WSR performance improvement obtained by the proposed precoders over other solutions available in the literature. For this study, transmit power of all the nodes is set to unity i.e., $P_b=P_u=P_r=1$. Also, $\sigma_r^2=\sigma^2 = 1$. The average per-hop SNR between BS $\leftrightarrow$ RS link is defined as \text{SNR}$^{(b)}= {h_b}^2={g_b}^2$. Similarly, average per-hop SNR between TUE $\rightarrow$ RS and RS $\rightarrow$ RUE is given as \text{SNR}$^{(u)}= {h_u}^2={g_u}^2$. Average WSR performance of the precoders is analysed for a) Balanced b) Unbalanced links. For balanced links, \text{SNR}$^{(b)}$ = \text{SNR}$^{(u)}$ = \text{SNR} are simultaneously varied from $0$ to $40$ dB. For unbalanced links, \text{SNR}$^{(b)}$ is fixed to $20$ dB as in \cite{relay_csi_sup_nw_code} and the \text{SNR}$^{(u)}$ is varied from $0$ to $40$ dB. For the sake of simplicity, downlink and uplink weights, $w_{u,m}$ and $w_{b,m}$, are set respectively as 1.5 and 0.5, $m=1,\cdots,M$, where $M$ is number of transmit streams. The average WSR performance is compared next for the following precoders:

1) \textbf{ZF precoder}: In \cite{eurasip_af_mimo_ref}, two precoders are proposed for symmetric TWR by adopting the interference mitigation approach. The first precoder is based on the ZF criterion and is designed to completely cancel the BI as well as inter-stream interference for the communicating nodes. The ZF precoder can be used in the asymmetric TWR scenario also, as it will lead to BS and RUE receiving the signal free from BI and inter-stream interference.
 
2) \textbf{MMSE precoder}: The second precoder in \cite{eurasip_af_mimo_ref} is designed using the MMSE criterion and is shown to have better performance than the ZF precoder.  It should be noted that the MMSE precoder does not cancel the BI and inter-stream interference completely. This residual BI can only be cancelled by the BS in asymmetric TWR, different from the symmetric case, where both the nodes can cancel the residual BI. The weighted sum-rate achieved by ZF and MMSE precoders is later maximized in \cite{eurasip_af_mimo_ref} by making an approximation to the mutual information values. The same procedure is used here while plotting the performance of these precoders. 

3) \textbf{Proposed precoder}: The precoder designed to cancel the BI and triangulate the MAC and BC phase channels at the relay in \eqref{dl_bpi_can_pre_ref} and \eqref{cp_pre_ref}, denoted as \textit{BI-cancelling-Channel-Triangularization} (BI-CT) precoder.  

\begin{figure}[htp]
  \begin{center}
   \iftoggle{SINGLE_COL}{\includegraphics[width=4.5in]{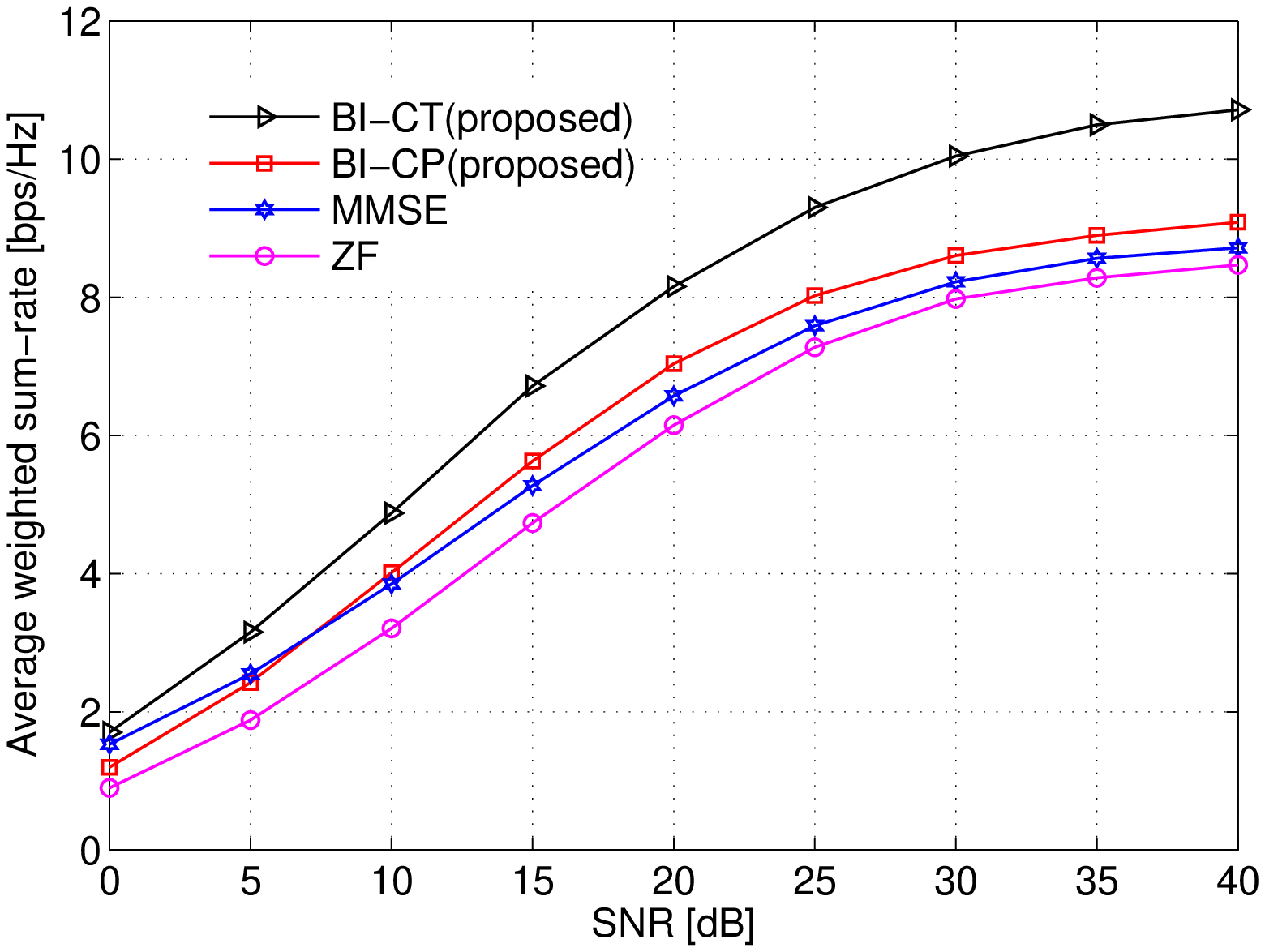}}{\includegraphics[width=3.55in]{wsr_af_4x2_unbal.eps}}
   \caption{\small Average WSR comparison for unbalanced links with $N=4$ antennas at the RS, $M=2$ antennas at the TUE, RUE and BS.} 
  \label{sum_rate_4x2_unbal_fig_ref}  
  \end{center}
    \end{figure}
In Fig. \ref{sum_rate_4x2_unbal_fig_ref}, the average WSR of different precoders are compared for the unbalanced links. Here, the performance of proposed baseline \textit{BI-cancelling-Channel-Parallelization} (BI-CP) precoder designed using block-ZF approach in \eqref{zf_prec_ref} is also plotted. It can be seen that the proposed BI-CT precoder outperforms all other precoders across all SNR values. Also, the proposed BI-CP precoder provides better average WSR than the ZF precoder at all SNRs and outperforms MMSE precoder at SNR $\ge$ 8 dB. The BI-CT and BI-CP precoders perform better than the other precoders due to the following reasons: 1) They are designed such that the BI is cancelled for RUE alone, whereas the ZF and MMSE precoders mitigate interference for the BS also; and 2) BI-CT precoder is a unitary precoder and avoids the channel matrix inversion unlike the BI-CP and ZF precoders. The channel-matrix inversion will lead to performance degradation if an ill-conditioned matrix has to be inverted. The penalty incurred due to channel inversion will be more pronounced as the number of antennas is increased at the nodes. This effect can be observed in Fig. \ref{sum_rate_8x4_unbal_fig_ref} where the number of antennas is doubled at each node when compared to the antenna configuration in Fig. \ref{sum_rate_4x2_unbal_fig_ref}. There is now a dramatic performance gap between the BI-CT precoder and the rest of the two precoders. BI-CT precoder provides 6 bps/Hz higher WSR than the BI-CP precoder at 30 dB (cf. Fig. \ref{sum_rate_8x4_unbal_fig_ref}) when compared to the improvement of 1.8 bps/Hz at same SNR in Fig. \ref{sum_rate_4x2_unbal_fig_ref}. Performance of BI-CP precoder is not included as its performance is only marginally better than the ZF and MMSE precoders.

\begin{figure}[htp]
  \begin{center}
  \iftoggle{SINGLE_COL}{\includegraphics[width=4.5in]{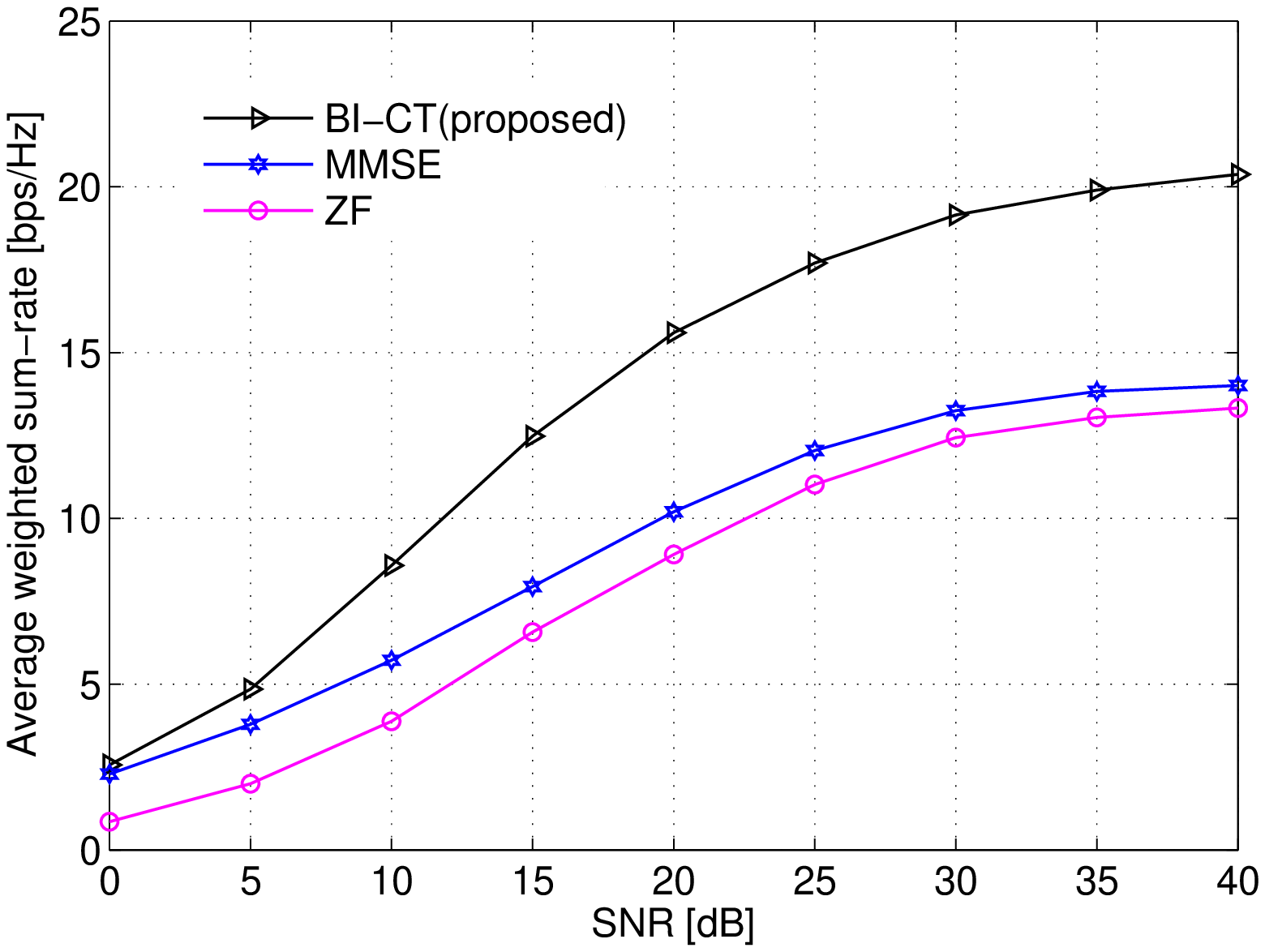}}{\includegraphics[width=3.55in]{wsr_af_8x4_unbal.eps}}
  \caption{\small Average WSR comparison for unbalanced links with $N=8$ antennas at the RS, $M=4$ antennas at the TUE, RUE and BS.}
  \label{sum_rate_8x4_unbal_fig_ref}  
  \end{center}
    \end{figure}

In Fig. \ref{sum_rate_8x4_bal_fig_ref}, performance of various precoders is compared for the balanced links. Here too, as expected, BI-CT performs better than the other precoders. 
\begin{figure}[htp]
  \begin{center}
    \iftoggle{SINGLE_COL}{\includegraphics[width=4.5in]{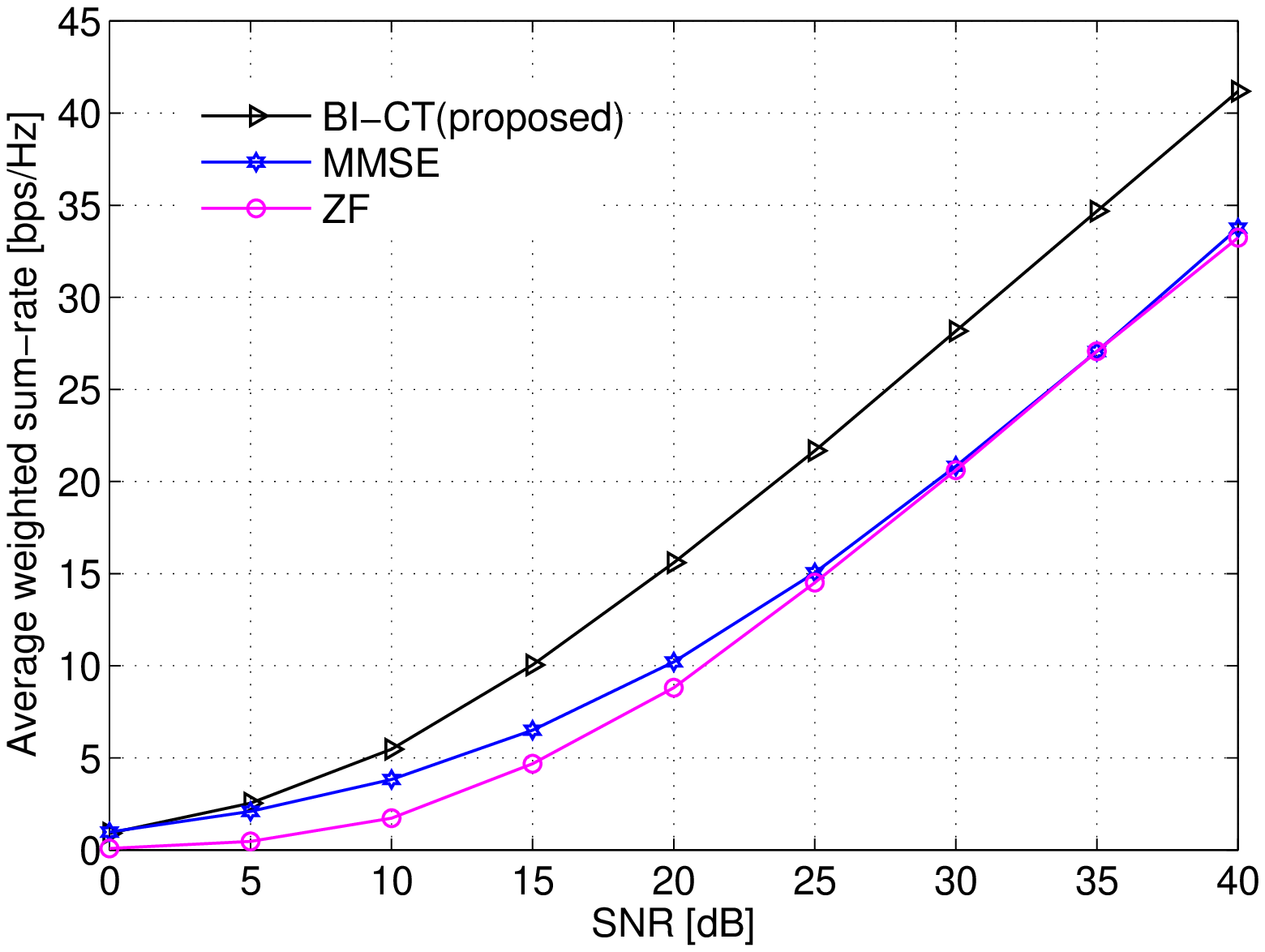}}{\includegraphics[width=3.55in]{wsr_af_8x4_bal.eps}}
    \caption{\small Average WSR comparison for balanced links with $N=8$ antennas at the RS, $M=4$ antennas at the TUE, RUE and BS.}
  \label{sum_rate_8x4_bal_fig_ref}  
  \end{center}
    \end{figure}
\subsection{WSR comparison of different transmission protocols in a cellular framework}
As shown in the previous section, proposed BI-CT precoder outperforms all other precoders with a considerable margin. In this section, performance of asymmetric TWR (ATWR) with BI-CT precoder is evaluated in a cellular framework and compared with the conventional one-way relaying and single-hop transmission. One-way relaying and single-hop transmission provide two other methods of information exchange between BS, TUE and RUE in the absence of proposed protocol. These performance comparisons will reveal the tangible performance gains provided by the ATWR over the other two options of data exchange.
 
1) \textbf{Optimal One-Way Relaying} (OWR): For OWR, we assume that a communication cycle consisting of a downlink phase and an uplink phase is divided into four time slots. The first two time slots are allocated for the downlink phase and the last two are used for the uplink phase. During the downlink phase, the relay  receives data from the BS in the first slot, performs non-regenerative linear processing and transmits it to the RUE during the second slot. During uplink phase, the relay will receive data from the TUE in the third slot and transmit this data (after non-regenerative linear processing) to the BS in the fourth slot. For OWR, separate precoders are required for the relay transmission during downlink and uplink phase. 

Let $\mathbf{W}_d$ be the relay precoder during the downlink phase. Let $\mathbf{H}_b$ and $\mathbf{G}_u$ be the channel matrices for BS$\rightarrow$RS and RS$\rightarrow$RUE links. If $\mathbf{U}_{b}\mathbf{\Delta}_{\mathbf{h}_{b}}\mathbf{U}_{b}^H$ and $\mathbf{V}_u\mathbf{\Delta}_{\mathbf{g}_u}\mathbf{V}_u^H$ are the eigenvalue decomposition \cite{hornandjohnson} of $\mathbf{H}_{b}\mathbf{H}_{b}^H$ and $\mathbf{G}_u^H\mathbf{G}_u$, respectively, then $\mathbf{W}_d = \mathbf{V}_u\mathbf{\Delta}_{u}\mathbf{U}_{b}^H$ is shown as the optimal precoder in \cite[(17)]{owr_opt_pre_ref},\cite{twr_opt_pre_ref7} to maximize the mutual information between BS and RUE.  Here $\mathbf{\Delta}_{u}$ is the diagonal power-allocation matrix. An algorithm to derive the optimal power allocation is also derived in \cite{owr_opt_pre_ref,twr_opt_pre_ref7}. We use this precoder to calculate the maximum end-to-end downlink rate observed by the RUE ($R_u$). The uplink precoder $\mathbf{W}_u$ and the corresponding end-to-end uplink rate observed by the BS ($R_b$) are also calculated in a similar fashion. WSR for OWR is then defined as $R_\textsf{sum} = \frac{1}{4}(w_uR_u + w_bR_b).$ The factor of $1/4$ is due to the fact that downlink and uplink phases are divided into four time slots. Similar to the last section, downlink and uplink weights, $w_u$ and $w_b$, are set as 1.5 and 0.5, respectively.

2) \textbf{Single-hop transmission} (Direct): For single-hop transmission, we assume that a communication cycle consisting of a downlink phase and an uplink phase is divided into two time slots. The first time slot is allocated for the downlink phase and the second slot is used for the uplink phase. If $\mathbf{H} \in \mathbb{C}^{M\times M}$ is the channel for the BS$\rightarrow$RUE link, the capacity of BS $\rightarrow$ RUE link is given as: $R_u = \log|\mathbf{I}_M+ \frac{P_b}{M\sigma^2}\mathbf{HH}^H |$ \cite{telatar_mimo_cap_ref}. Here we assume that the CSI is available only at the RUE and not at the BS, consistent with the asymmetric TWR model. Similarly, the capacity of TUE$\rightarrow$BS link with the CSI available at the BS is given as: $R_b = \log|\mathbf{I}_M+ \frac{P_u}{M\sigma^2}\mathbf{GG}^H |$, where $\mathbf{G} \in \mathbb{C}^{M\times M}$ is the channel for the TUE$\rightarrow$BS link. The elements of uplink and downlink channels, $\mathbf {H} \text{ and } \mathbf{G}$, are independent and are distributed as $\mathcal{CN}(0,h^2)$  and  $\mathcal{CN}(0,g^2)$, respectively. WSR for direct transmission is then calculated as  $R_\textsf{sum} = \frac{1}{2}(w_uR_u + w_bR_b)$. The factor of $1/2$ is due to the fact that downlink and uplink phases are divided into two time slots. Here also downlink and uplink weights, $w_u$ and $w_b$, are set as 1.5 and 0.5, respectively. 

The system parameters used for comparing the performance of the above three modes of information exchange are listed in Table \ref{table1}. For the fair evaluation of different transmission options, RS transmit power is added to the BS transmit power for the single-hop transmission. The WSR is obtained by employing the precoder on a single subcarrier of an orthogonal frequency division multiplexing (OFDM) based cellular system. Transmit power of the nodes is therefore normalized to obtain per Hz transmission power.  

\begin{table}[!ht]
\renewcommand{\arraystretch}{1.15}
\centering
\begin{tabular}{|l|l|}  \cline{1-2}  
Carrier Frequency  & 2 GHz   \\    \cline{1-2}
Thermal Noise & -174 dBm/Hz \\ \cline{1-2}
System Bandwidth & 10 MHz \\ \cline{1-2}
Noise Figure & 7 dB \\ \cline{1-2}
\iftoggle{SINGLE_COL}{
BS Transmit power & 46 dBm \\    \cline{1-2}
UE Transmit power & 24 dBm \\    \cline{1-2}
}
{BS/UE Transmit power & 46 dBm/24 dBm \\    \cline{1-2}}
BS/RS/UE height  & 30m/15m/1m  \\    \cline{1-2}
BS-RS distance   & 1 Km \\ \cline{1-2}
BS-RS channel model   & IEEE 802.16j, Type D\cite{16j_ch_mod_ref} \\ \cline{1-2}
\iftoggle{SINGLE_COL}{\end{tabular}
\begin{tabular}{|l|l|}  \cline{1-2}}{}  
\multicolumn{2}{|c|}{ \textbf{Coverage-extension parameters}} \\ \cline{1-2}
RS-MS channel model   & IEEE 802.16j, Type B\cite{16j_ch_mod_ref} \\ \cline{1-2}
BS-MS channel model   & IEEE 802.16j, Type B\\ \cline{1-2}
RS Transmit power  & 39 dBm  \\    \cline{1-2}
\multicolumn{2}{|c|}{\textbf{Coverage-hole parameters}} \\ \cline{1-2}
RS-MS channel model   & IEEE 802.16j, Type E\cite{16j_ch_mod_ref} \\ \cline{1-2}
BS-MS channel model   & IEEE 802.16j, Type E\\ \cline{1-2}
RS Transmit power  & 30 dBm  \\    \cline{1-2}
Penetration loss & 10 dB	\\ \cline{1-2}
\end{tabular}
\caption{System parameters}
\label{table1}
\end{table}
Among other scenarios, the deployment of  infrastructure relays is envisaged in \cite{infra_relay_ref1,infra_relay_ref2} for: 1) Enhancing coverage in the areas where capacity of direct links between BS and UEs is low due to high path loss. Such areas can exist at the cell edge \cite{gp_owr_ref,infra_relay_ref1}; and 2) Providing coverage in the areas where capacity of direct link is nearly zero e.g., a coverage hole. We limit our study to these coverage-oriented scenarios in this section. The placement of relays in these scenarios is such that they are likely to cause minimal inter-cell interference. Further, it is also assumed that the low inter-cell interference can be handled using concepts like scheduling, fractional frequency reuse \cite{relay_lte_ref_fabian}. We therefore concentrate on a single cell framework with a BS, RS and two UEs.  
\begin{figure}[htp]
  \begin{center}
    \iftoggle{SINGLE_COL}{\includegraphics[width=4.5in]{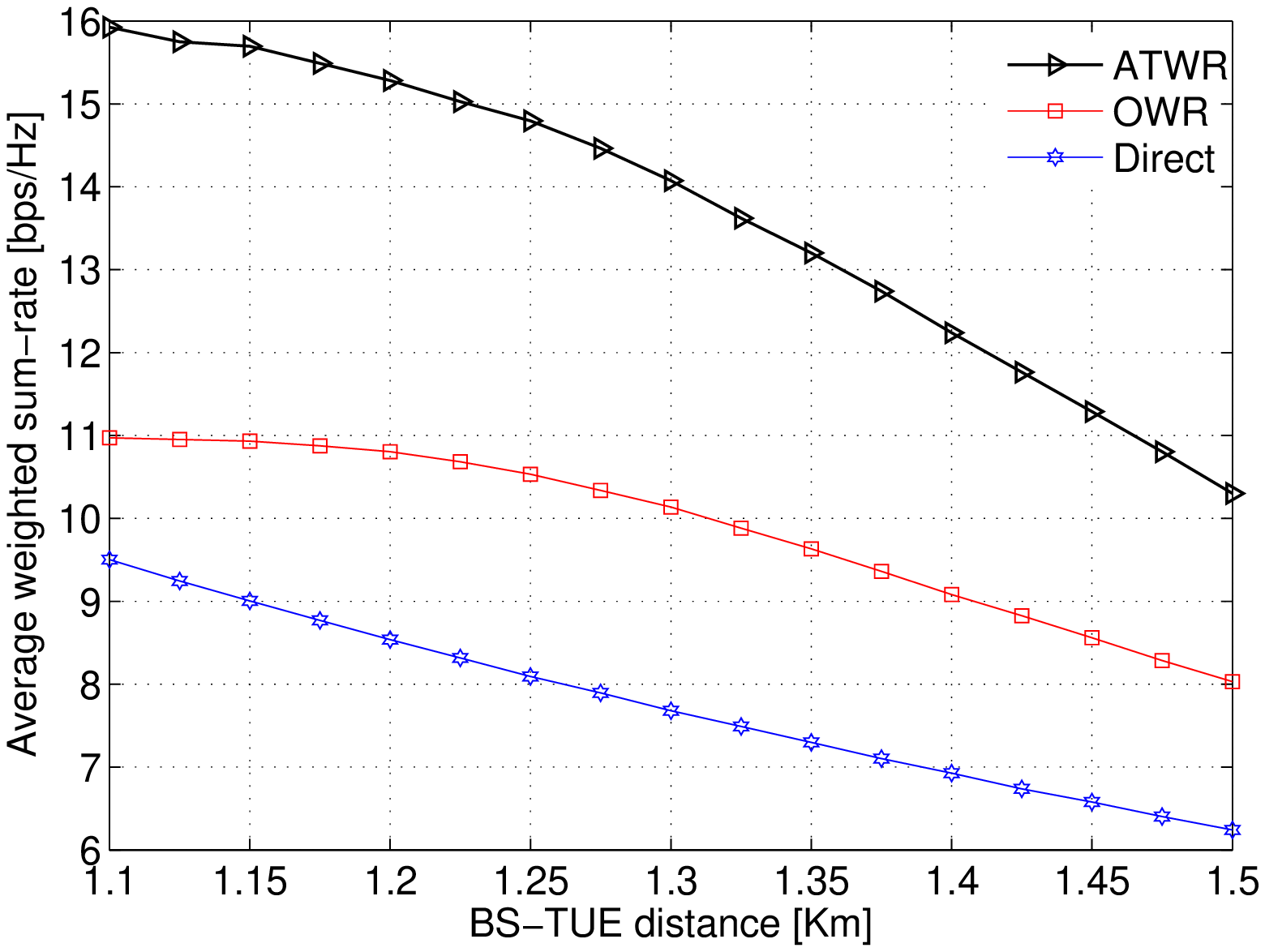}}{\includegraphics[width=3.55in]{wsr_af_4x2_cov_ext_39dB.eps}}
  \caption{\small Average WSR comparison for coverage-extension scenario with $N=4$ antennas at the RS, $M=2$ antennas at the TUE, RUE and BS. Here BS-RUE distance = 1.5 Km.}
  \label{sum_rate_4x2_cov_ext_hp_fig_ref}  
  \end{center}
    \end{figure}

As mentioned in the Table \ref{table1}, the RS is located at a fixed distance of 1 Km from the BS. For the coverage-extension scenario, we consider a site of radius 500m around the RS where coverage needs to be provided by the RS. For this study, location of RUE is fixed at the edge of the RS site i.e., a distance of 500m from the relay and TUE-RS distance is varied from 100m to 500m. In Fig. \ref{sum_rate_4x2_cov_ext_hp_fig_ref}, where WSR curves are plotted, it can be seen that the ATWR provides significantly higher WSR than the OWR and the baseline direct-transmission across the entire range of distance of operation. At a BS-TUE distance of 1.3 Km (equivalent RS-TUE distance of 0.3 Km), there is a difference of $\sim$ 4 bps/Hz in the WSR performance of ATWR and OWR.

For the coverage-hole scenario, a site of radius of 100m is considered around the RS where the coverage-hole needs to be plugged. Here RUE is located at a fixed distance of 50m from the relay while TUE-RS distance is varied from 10m to 100m. In Fig. \ref{sum_rate_4x2_cov_hole_fig_ref}, where the ATWR performance is compared with the OWR and the direct transmission, it is clear that the ATWR provides much better WSR than the OWR through out the distance of operation. The  capacity of direct transmission in a coverage-hole is negligible when compared to the ATWR. 
\begin{figure}[htp]
  \begin{center}
    \iftoggle{SINGLE_COL}{\includegraphics[width=4.5in]{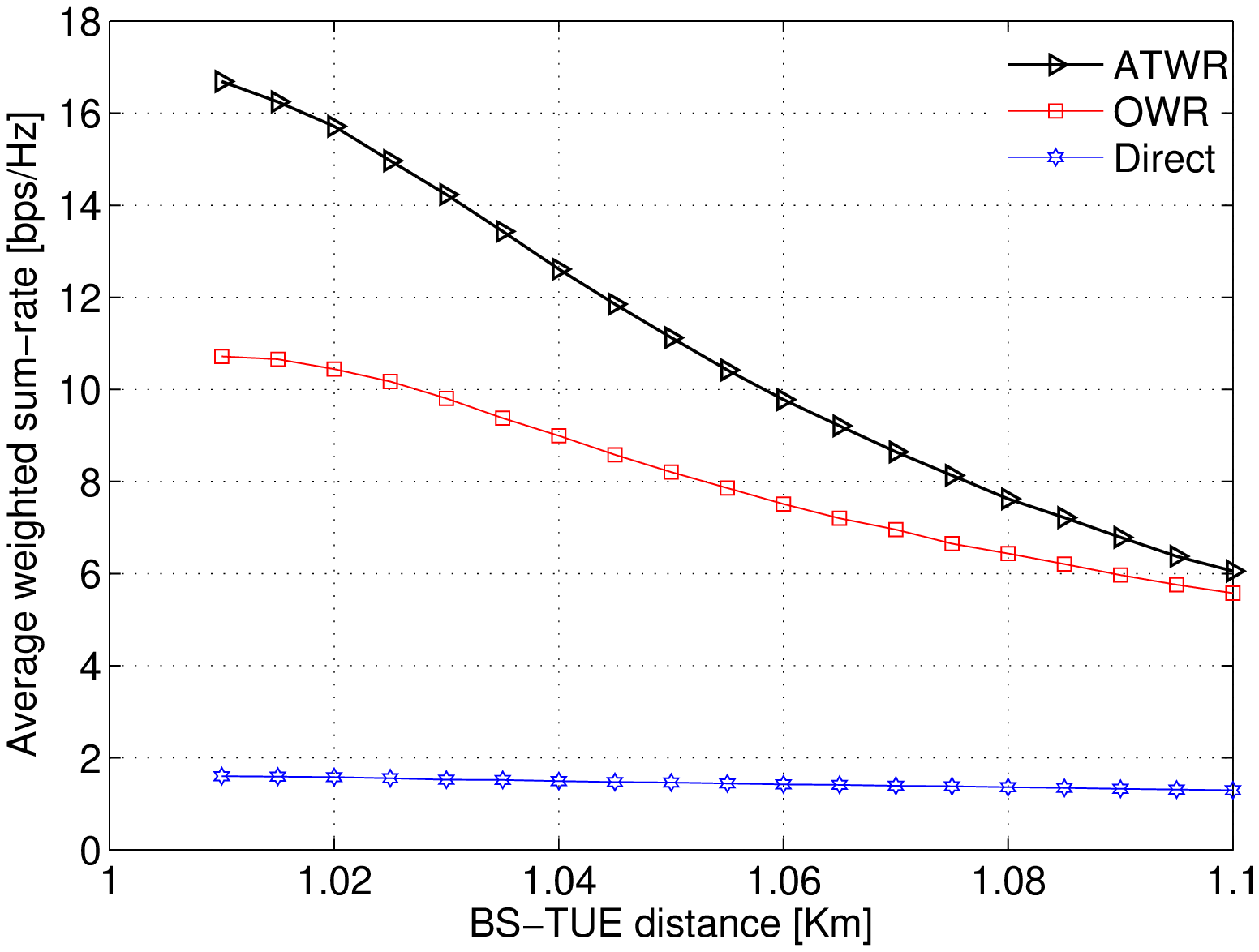}}{\includegraphics[width=3.55in]{wsr_af_4x2_cov_hole.eps}}
  \caption{\small Average WSR comparison for coverage-hole scenario with $N=4$ antennas at the RS, $M=2$ antennas at the TUE, RUE and BS. Here BS-RUE distance = 1.05 Km.}
  \label{sum_rate_4x2_cov_hole_fig_ref}  
  \end{center}
    \end{figure}
\section{Conclusion}
\label{conclusion_sec_ref}
The assumption of simultaneous exchange of data traffic in conventional TWR is generally not applicable to cellular systems. This paper has considered the problem of asymmetric TWR and has proposed a new protocol to handle the non-simultaneous data exchange. Due to the back-propagating interference (BI) observed by the receiving UE (RUE) in the asymmetric TWR, communication between three nodes is possible either by doubling the number of RUE antennas at the RUE or by sacrificing the spatial resources. We have designed a novel linear precoder at the relay to completely cancel the asymmetric BI. Consequently, there is no need to increase the number of RUE antennas or sacrifice the spatial resources. The structure of the proposed precoder is  exploited to triangulate the MAC and BC phase channels of BS and RUE, thus simplifying their receiver design. Due to channel triangularization, the weighted sum-rate (WSR) maximization reduces to power allocation problem, and can be cast as a geometric program in the high-SNR regime. With the WSR maximization, it is possible for the relay to assign individual priorities to each stream to satisfy their quality-of-service constraints. As a byproduct of WSR maximization, the solution of relay power minimization under given QoS constraints is also provided. The WSR of the proposed precoders is compared with the state-of-the-art precoders for different antenna configurations via simulations. The results indicate that the WSR of the proposed precoder outperforms the conventional ZF and MMSE precoders at all values of SNR by a significant margin. The salutary performance benefits of the asymmetric two-way relaying over conventional one-way relaying and single-hop transmission are demonstrated in two different coverage-limited cellular scenarios.

\appendices
\numberwithin{equation}{section}
\section{Non-convexity of the WSR maximization problem.}
\label{non_con_app_ref}
For the sake of brevity, SNR observed by the $\widetilde m$th stream of RUE and BS for the designed precoder is expressed as
\begin{equation}
\text{SNR}_{i,\widetilde{m}} = \frac{a_{\widetilde m}\delta_{i,m}}{\sigma_r^2(\sum_{j=1}^Mb_{i,j}^{\widetilde m}\delta_{u,j} + c_{i,j}^{\widetilde m}\delta_{b,j}) + \sigma^2}.
\label{gen_snr_eq_ref}
\end{equation}
Recall that $\widetilde{m}= M - m +1$ and $i \in \{u,b\}$. The exact coefficients $\{a_{\widetilde m}, b_{i,j}^{\widetilde m}, c_{i,j}^{\widetilde m}\} \ge 0$ are given in \eqref{bs_ue_fin_snr_sic} for the designed precoder. The objective function in \eqref{af_opt_prob_def} can therefore be re-written as: 
\begin{multline}
 \sum_{\forall i}\Bigg\{\sum_{\forall m} {w_{i,m}}\log\Big(\sigma_r^2\Big\{{\sum_{j=1}^Mb_{i,j}^{\widetilde m}\delta_{u,j} + c_{i,j}^{\widetilde m}\delta_{b,j}}\Big\} +  \sigma^2+ \\a_{\widetilde m}\delta_{i,m}\Big)
 -\sum_{\forall m} {w_{i,m}}\log\Big(\sigma_r^2\Big\{\sum_{j=1}^Mb_{i,j}^{\widetilde m}\delta_{u,j} + c_{i,j}^{\widetilde m}\delta_{b,j}\Big\} + \sigma^2 \Big)\Bigg\}\nonumber. 
\end{multline}
It can be seen that the objective function is a difference of two concave functions of the variables $\delta_{u,j}$ and $\delta_{b,j}, \ j = 1 \cdots M$ and is therefore non-convex.
\section{Proof of Lemma \ref{pow_const_lemma_ref_sic}}
\label{lq_qr_app_ref}
In this appendix, we show that the power constraint in the optimization problem in \eqref{af_opt_prob_def} can be expressed as a posynomial. From \eqref{pre_def_eq_ref}, the precoder $\mathbf{W}$ can be decomposed as $\mathbf{W}=\mathbf{MDF}$. Channel triangularization precoder matrix $\mathbf{D}$ (cf. \eqref{d_mat_def_ref} and \eqref{isi_pre_eq_ref}) can be re-written as
\begin{align}
\mathbf{D}=
\underbrace{\left[\begin{array}{ccrr} \mathbf{\Pi}_u & \mathbf{0} \\\mathbf{0}  &\mathbf{\Pi}_b  \end{array}\right]}_{\mathbf{\Pi}} 
\underbrace{\left[\begin{array}{ccrr} \mathbf{0} & \mathbf{\Delta}_u \\\mathbf{\Delta}_b  &\mathbf{0}  \end{array}\right]}_{\mathbf{\Delta}}  
\underbrace{\left[\begin{array}{ccrr} \mathbf{\Theta}_b & \mathbf{0} \\\mathbf{0}  &\mathbf{\Theta}_u  \end{array}\right]}_{\mathbf{\Theta}}  
\label{d_mat_decom_eq_ref}
\end{align}
Note that $\mathbf{\Pi}$ and $\mathbf{\Theta}$ are unitary matrices and $\mathbf{\Delta}$ is anti-diagonal matrix. The precoder $\mathbf{W}$ can now be re-expressed  as $\mathbf{W}=\mathbf{M\Pi \Delta \Theta F}=\bar{\mathbf{M}}\mathbf{\Delta}\bar{\mathbf{F}}$, where $\bar{\mathbf{M}} = \mathbf{M\Pi}$ and $\bar{\mathbf{F}} = \mathbf{\Theta F}$. The unitary structure of $\mathbf{\Pi}$ ensures that $\bar{\mathbf{M}}$ has orthonormal columns while unitary  $\mathbf{\Theta}$ ensures orthonormal rows for $\bar{\mathbf{F}}$. The power constraint in \eqref{relay_pow_const_eq_ref} is next simplified to show that it can be expressed as a posynomial. 
\begin{align}
P_r &\ge \text{Tr}\left(\mathbf{WHQ}\mathbf{H}^H\mathbf{W}^H +\sigma_r^2\mathbf{W}\mathbf{W}^H\right) \\
&=\sum_{j=1}^M\left\{\rho_u\|\mathbf{Wh}_{j}^u\|^2 + \rho_b\|\mathbf{Wh}_{j}^b\|^2 \right\}+ \sigma_r^2 \text{ Tr}(\mathbf{WW}^H)\nonumber\\
 &=\sum_{j=1}^M \rho_u\|\mathbf{\bar M\Delta \bar Fh}_{j}^u\|^2 + \rho_b\|\mathbf{\bar M\Delta \bar Fh}_{j}^b\|^2 + \sigma_r^2\text{Tr}(\mathbf{WW}^H) \nonumber \\
 &=\sum_{j=1}^M \rho_u\|\mathbf{\bar M \Delta q}_{j}^u\|^2 + \rho_b\|\mathbf{\bar M \Delta q}_{j}^b\|^2 + \sigma_r^2 \text{Tr}(\mathbf{WW}^H) \nonumber \\
 &\stackrel{(a)}{=}\sum_{j=1}^M \rho_u\|\mathbf{\Delta q}_{j}^u\|^2 + \rho_b\|\mathbf{\Delta q}_{j}^b\|^2 + \sigma_r^2 \text{Tr}(\mathbf{WW}^H) \nonumber \\
 &\stackrel{(b)}{=}\sum_{j=1}^M \rho_u\|\mathbf{\Delta q}_{j}^u\|^2 + \rho_b\|\mathbf{\Delta q}_{j}^b\|^2 + \sigma_r^2 \text{Tr}(\mathbf{\Delta\Delta}^H) \nonumber\\
 &= \sum_{m=1}^{M}\sum_{j=1}^{M}\bigg(\big\{\rho_u|\mathbf{q}_{j,\widehat{m}}^u|^2 + \rho_b|\mathbf{q}_{j,\widehat{m}}^b|^2 + \sigma_r^2\big\}\delta_{u,m} 
 \iftoggle{SINGLE_COL}{}
 {\nonumber\\
  &\ \ \ }
  + \big\{\rho_u|\mathbf{q}_{j,\widetilde{m}}^u|^2 + \rho_b|\mathbf{q}_{j,\widetilde{m}}^b|^2 + \sigma_r^2\big\}\delta_{b,m}\bigg) \label{sic_posy_eq_ref}
 \end{align}
Here $\mathbf{h}_{j}^u$ and $\mathbf{h}_{j}^b$ denote the $j^{th}$ column of $\mathbf{H}_u$ and $\mathbf{H}_b$, respectively. Also, $\mathbf{q}_j^u = \mathbf{\bar Fh}_j^u = [q_{1,j}^u, \cdots, q_{2M,j}^u]^T $ and $\mathbf{q}_j^b = \mathbf{\bar Fh}_j^b = [q_{1,j}^b, \cdots, q_{2M,j}^b]^T $.
Also $\widehat m = 2M - m +1$ and $\widetilde m = M - m +1$. In $(a)$ we have used the fact that $\mathbf{\bar M}$ has orthonormal columns by design. Equality in $(b)$ can be derived by using the following facts: 1) for any arbitrary matrices $\mathbf{A, B}$ of compatible dimensions, Tr($\mathbf{AB}$) = Tr($\mathbf{BA}$); and 2) $\mathbf{\bar F}$ has orthonormal rows and $\mathbf{\bar M}$ has orthonormal columns. It can be seen that all the coefficients of $\delta_{u,m}$ and $\delta_{b,m} , \forall m$, are non-negative. \eqref{sic_posy_eq_ref} is a valid posynomial.
\section{Generalized GP as an equivalent GP}
\label{gen_posy_app_ref}
Towards this end, we first express the optimization problem in \eqref{final_af_opt_prob_ref} in the epigraph form \cite{boyd_book} i.e.,
\begin{equation}
\begin{aligned}
 & \underset{\boldsymbol{\delta}\succeq 0} {\text{ Min. }} 
&& t\\
&\text{ s.t. } 
&&\prod_{m=1}^M (f_{u,m}(\boldsymbol{\delta}))^{w_{u,m}}(f_{b,m}(\boldsymbol{\delta}))^{w_{b,m}} \le t \text{ and } \eqref{relay_pow_const_eq_ref},
\label{ggp_con_ref0}
\end{aligned}
\end{equation}
where $f_{u,m}(\boldsymbol{\delta}) = \text{ISNR}_{u,m}(\boldsymbol{\delta})$ and $f_{b,m}(\boldsymbol{\delta}) = \text{ISNR}_{b,m}(\boldsymbol{\delta})$. The generalized posynomial in the objective function is transformed into a generalized posynomial constraint (GPC). We next show that the GPC in \eqref{ggp_con_ref0} can be transformed into equivalent posynomial constraint (PC). By using the auxiliary variables $(t_{u,m}, t_{b,m}), m=1,\cdots, M$, the GPC can be re-expressed as 
\begin{equation}
\begin{aligned}
 &\prod_{m=1}^M ({t_{u,m}})^{w_{u,m}}({t_{b,m}})^{w_{b,m}} \le t, \\
 &f_{u,m}(\boldsymbol{\delta})\le {t_{u,m}}\text{ and } f_{b,m}(\boldsymbol{\delta})\le{t_{b,m}}, \forall m 
\label{ggp_con_ref1}
 \end{aligned}
 \end{equation}
Note that the $2M+1$ constraints as expressed in \eqref{ggp_con_ref1} are valid PC. We next show that the GPC in \eqref{ggp_con_ref0} and the PC in \eqref{ggp_con_ref1} are equivalent. Let $t,t_{u,m},t_{b,m} \text{ and } \boldsymbol{\delta}$ satisfy \eqref{ggp_con_ref1}. Since the GPC in \eqref{ggp_con_ref0} is monotonically non-decreasing in each of its argument (due to positive weights), it implies that GPC holds. Conversely, if the GPC holds in \eqref{ggp_con_ref0}, then by assigning ${t_{u,m}} = f_{u,m}(\boldsymbol{\delta}), {t_{b,m}} = f_{b,m}(\boldsymbol{\delta}), \forall m$, we observe that $\prod_{m=1}^M ({t_{u,m}})^{w_{u,m}}({t_{b,m}})^{w_{b,m}} \le t$, $f_{u,m}(\boldsymbol{\delta})= {t_{u,m}} \text{ and } f_{b,m}(\boldsymbol{\delta}) = {t_{b,m}}$. This implies that \eqref{ggp_con_ref1} is satisfied. The GPC can thus be expressed as equivalent PC and the GGP can be solved as a GP. Note that the  power constraint in \eqref{relay_pow_const_eq_ref} is a posynomial as shown in appendix \ref{lq_qr_app_ref}. 


\end{document}